\documentclass[twocolumn,superscriptaddress,preprintnumbers,amsmath,amssymb]{revtex4}

\usepackage{graphicx,amsmath,dcolumn,bm}
\usepackage{dcolumn}% Align table columns on decimal point

\begin{document}

\newcommand{\BQ}{\begin{equation}}
\newcommand{\EQ}{\end{equation}}
\newcommand{\BQA}{\begin{eqnarray}}
\newcommand{\EQA}{\end{eqnarray}}
\newcommand{\be}{\begin{eqnarray}}
\newcommand{\ee}{\end{eqnarray}}
\newcommand{\NN}{\nonumber \\}
\newcommand{\del}{\partial}
\newcommand{\tr}{{\rm tr}}
\newcommand{\Tr}{{\rm Tr}}
\newcommand{\Path}{{\rm P}\,}
\newcommand{\ket}[1]{\left.\left\vert #1 \right. \right\rangle}
\newcommand{\bra}[1]{\left\langle\left. #1 \right\vert\right.}
\newcommand{\ketrm}[1]{\vert {\rm #1} \rangle}  % for |phys>
\newcommand{\brarm}[1]{\langle {\rm #1} \vert}  % for <phys|
\newcommand{\V}{\widetilde V}
\newcommand{\U}{\widetilde U}
\newcommand{\e}{{\rm e}}
\def\simge{\mathrel{%
   \rlap{\raise 0.511ex \hbox{$>$}}{\lower 0.511ex \hbox{$\sim$}}}}
\def\simle{\mathrel{
   \rlap{\raise 0.511ex \hbox{$<$}}{\lower 0.511ex \hbox{$\sim$}}}}

\vspace*{2mm}
\title{Shear viscosity of a hadronic gas mixture}

\author{Kazunori Itakura}
\affiliation{High Energy Accelerator Research Organization (KEK),
          Oho 1-1, Tsukuba, Ibaraki, 305-0801, Japan}

\author{Osamu Morimatsu}
\affiliation{High Energy Accelerator Research Organization (KEK),
          Oho 1-1, Tsukuba, Ibaraki, 305-0801, Japan}
\affiliation{Department of Physics, University of Tokyo, 7-3-1 
Hongo Bunkyo-ku Tokyo 113-0033, Japan}

\author{Hiroshi Otomo}
\affiliation{Department of Physics, University of Tokyo, 7-3-1 
Hongo Bunkyo-ku Tokyo 113-0033, Japan}

\date{November 7, 2007}

\begin{abstract}
We discuss in detail the shear viscosity coefficient $\eta$ and 
the viscosity to entropy density ratio $\eta / s$ of a hadronic 
gas comprised of pions and nucleons. In particular, we study the effects of 
baryon chemical potential  on $\eta$ and $\eta/s$. 
We solve the relativistic quantum Boltzmann equations with 
binary collisions ($\pi\pi$, $\pi N,$ and $NN$) 
for a state slightly 
deviated from thermal equilibrium at temperature $T$ and 
baryon chemical potential $\mu$.  The use of phenomenological 
amplitudes in the collision terms, which are constructed to 
reproduce experimental data, greatly helps to extend the validity 
region in the $T$-$\mu$ plane. 
The total viscosity coefficient $\eta(T,\mu)=\eta^\pi+\eta^N$ 
increases as a function of $T$ and $\mu$, 
indirectly reflecting energy dependences of binary cross sections. 
The increase in $\mu$ direction is due to enhancement of the 
nucleon contribution $\eta^N$ while the pion contribution $\eta^\pi$
diminishes with increasing $\mu$. On the other hand, 
due to rapid growth of entropy density, the ratio $\eta/s$ becomes 
a decreasing function of $T$ and $\mu$ in a wide region of the
$T$-$\mu$ plane. In the kinematical region we investigated 
$T<180$~MeV, $\mu<1$~GeV, the smallest value of $\eta/s$ is 
about 0.3. 
Thus, it never violates the conjectured lower bound 
$\eta/s= 1/4\pi\sim 0.1$. The smallness of $\eta/s$
in the hadronic phase and its continuity at $T\simeq T_c$
(at least for crossover at small $\mu$) implies that the ratio 
will be small enough in the deconfined phase $T\simge T_c$.
There is a nontrivial structure at low temperature and at around 
normal nuclear density. We examine its possible 
interpretation as the liquid-gas phase transition.

\end{abstract}
%\pacs{Valid PACS appear here}

\maketitle

%%%%%%%%%%%%%%%%%%%%%%%%%%%%%%%%%%%%%%%%%%%%%%%%%%%%%
%%%%%%%%%%%%%%%%%%  INTRODUCTION  %%%%%%%%%%%%%%%%%%%
%%%%%%%%%%%%%%%%%%%%%%%%%%%%%%%%%%%%%%%%%%%%%%%%%%%%%

\section{Introduction}

Shear viscosity of a hot QCD matter has been attracting 
much attention in recent years. The major reason for that
is the intriguing experimental discovery 
that the matter created in heavy-ion collisions at Relativistic 
Heavy Ion Collider (RHIC) in Brookhaven National Laboratory 
could be close to a perfect fluid \cite{RHIC}. 
This unexpected result has driven people to think about 
{\it strongly-interacting} quark-gluon plasma (abbreviated as ``sQGP")
which may be realized at temperature just above the critical 
temperature $T\simge T_c$. 
As a result of strong-coupling nature, sQGP is thought 
to have very small shear viscosity, which is however not 
directly confirmed yet in a satisfactory way. 
In fact, only a few things are understood about sQGP 
because we will not be able to investigate it 
within standard perturbative QCD techniques. 
The only technique available now 
(except for lattice simulations which are not analytic methods) 
is the one based on the AdS/CFT correspondence, which 
relates strongly-coupled supersymmetric Yang-Mills theories 
to weakly-coupled gravity theories. 

There is an interesting outcome from the AdS/CFT analysis
in relation to the shear viscosity: It has been conjectured 
that there would be a lower bound in the ``shear viscosity 
coefficient to the entropy density ratio" (or simply the 
``viscosity to entropy ratio") $\eta/s\ge 1/4\pi$ \cite{KSS}. 
We call this ``the KSS bound" after authors' names of 
Ref.~\cite{KSS}. The lowest value $\eta/s=1/4\pi$ is satisfied 
by several super Yang-Mills theories in the large $N_c$ limit 
(strong coupling limit), which suggests that the bound could 
be universal. Of course, there is no guarantee for this bound 
to hold in real (non-supersymmetric) QCD whose gravity dual 
is not found, but interestingly enough, the values of $\eta/s$ 
extracted from RHIC experiments \cite{Gavin} and from lattice 
simulations \cite{Nakamura,Meyer} seem to be small enough and 
close to the lower bound. 

On the other hand, there is an important empirical observation 
which can be seen in many substances such as helium, nitrogen, 
and water: The ratio $\eta/s$ has a minimum at or 
near the critical temperature \cite{Kapusta} (see also \cite{Hirano}). 
More precisely, the ratio shows a cusp at $T_c$ for the first 
order transition, while it has a convex shape for the crossover  
with its bottom around the (pseudo) critical temperature. 
Since this behavior is observed in many substances, it is 
expected to be universal. 
Recall that the phase transition in QCD is most probably 
crossover at least for low densities. 
Therefore, what we naturally expect for the QCD matter from the 
two observations mentioned above is that the ratio $\eta/s$ will 
have the minimum at $T\sim T_c$, 
and the numerical value at that point will be close to the KSS bound 
$\eta/s\sim 0.1$. It is of primary importance to check 
whether this expectation is indeed the case or not, 
and to understand the properties of the QCD matter around $T_c$
not only for $T\simge T_c$.
These considerations motivated us to investigate the shear viscosity 
in QCD {\it from the hadronic phase} $T\simle T_c$. 
Notice that we can indirectly 
study the properties of sQGP from below $T_c$ because physical 
quantities such as the ratio $\eta/s$ will be continuous at $T_c$ 
for the crossover transition. 

Transport properties of a meson gas have been studied 
by several people. Many of the calculations are based on 
the Boltzmann equations with the Chapman-Enskog 
method which is a standard approach for weak dissipative phenomena,
especially for computing  transport coefficients \cite{deGroot}. 
Differences among several papers \cite{Gavin2,Prakash,Davesne,Dobado1,Dobado2}
include kinetic or statistical properties of particles 
(relativistic or nonrelativistic, quantum or classical),
species of mesons (pions, kaons, etc.), 
and the cross sections in the collision terms. 
For example, Ref.~\cite{Dobado1} treated a nonrelativistic 
quantum pion gas with the binary cross section given by the 
leading order chiral perturbation theory (LO-ChPT). 
Recently, similar problems have been revisited in relation 
to the KSS bound \cite{Chen1,Dobado3}.
In Ref.~\cite{Chen1}, the ratio $\eta/s$ computed with the cross
section in LO-ChPT turned out to violate the KSS bound 
for temperature beyond  $T_c\sim 170$~MeV, 
and it was speculated that 
such violation could be related to the existence of phase transition. 
However, soon after that, it was shown in Ref.~\cite{Dobado3} 
that the KSS bound is not violated in a pionic gas if one computes 
the shear viscosity with a phenomenological cross section using 
the experimental phase shifts. What we have learned from these 
papers is the following: (i) we have to be careful when we use 
the cross section from effective field theories, and (ii) the 
ratio $\eta/s$ of a relativistic pion gas is small enough at 
relatively large temperature $T\simle T_c$, but does not violate 
the KSS bound $\eta/s\ge 1/4\pi$.

At this point, there comes a natural question: how does $\eta/s$
change if one adds nucleons to the pure pion gas? 
Naively, we expect that the ratio $\eta/s$ will {\it decrease} as 
number of nucleons is increased 
because the pion cross section will effectively enhance in 
the presence of nucleons, yielding smaller shear viscosity, 
while the entropy will increase. Thus, there is a chance that 
the ratio could violate the KSS bound if the bound does not change. 
Notice that the pion-nucleon gas is the minimum requisite which 
allows us to study the effects of {\it baryon chemical potential} 
$\mu$.
Therefore, it is quite interesting and important to investigate 
the $\mu$ dependence of $\eta/s$ in the pion-nucleon gas. 
Such investigation will also urge people to study the
(possible) $\mu$-dependence 
of the KSS bound. In fact, the ``universality'' of the KSS 
bound has not been tested at finite baryon chemical potential. 

Most recently, the authors of Ref.~\cite{Chen1} have applied 
their framework to the pion-nucleon 
gas to study the behavior of $\eta/s$ in the $T$-$\mu$ 
plane \cite{Chen2}. However, their focus was not on the KSS bound 
but on the new finding: a valley structure in $\eta/s$ 
at low temperature and large chemical potential which they 
argued would correspond to the nuclear liquid-gas 
phase transition. Although this is a very interesting 
suggestion, their results should be critically checked 
since they are based on the effective field theories 
whose validity region is severely limited. On the other hand, 
a realistic calculation of the viscosity in the pion-nucleon gas
was performed some time ago by Prakash et al. \cite{Prakash}. 
Remarkably, they used the binary cross sections in the collision 
terms, which roughly reproduce experimental data. 
However, unfortunately, the dependence 
on baryon chemical potential was not investigated in detail. 
Besides, this calculation is based on {\it classical} 
Boltzmann equations, and thus cannot be 
applied to relatively large chemical potential where the effects 
of Fermi statistics is expected to be large. 

In view of the present situation mentioned above,  
what we should do is rather evident: for the purpose of 
studying the $\mu$ dependence of the shear viscosity $\eta$
and the ratio $\eta/s$ in a pion-nucleon gas, we treat the 
relativistic quantum Boltzmann equations with binary cross sections
which are determined to reproduce experimental data. 
We are very careful about the range of validity of our framework.
We also check whether the valley structure found 
in Ref.~\cite{Chen2} indeed exists even with the phenomenological
cross sections. 
It is also important to compare our results with those from 
hadron cascade simulations. For example, the shear viscosity 
coefficient is computed for a meson gas in Ref.~\cite{Muronga}
and for a meson-baryon gas in Ref.~\cite{Muroya}.

The paper is organized as follows: in the next section, 
we explain the relativistic quantum Boltzmann equations 
for a dilute pion-nucleon gas, and define the cross sections we 
use in the collision terms. 
We treat the small deviation from the thermal equilibrium 
to the linear order (the Chapman-Enskog method), and give 
the shear viscosity coefficient $\eta$ through the solutions 
to the Boltzmann equations. In Sect.~III, we present 
our numerical results for a pure pion gas and a pion-nucleon gas mixture. 
We introduce a criterion which measures the validity region of the 
calculations. The use of phenomenological 
cross sections is very important to enlarge the range of validity. 
We discuss in detail the effects of chemical potential on 
$\eta$ and $\eta/s$ and examine the interpretation of 
the valley structure as the liquid-gas phase transition.
 Summary is given in the last section.

%%%%%%%%%%%%%%%%%%%%%%%%%%%%%%%%%%%%%%%%%%%%%%%%%%%%%
%%%%%%%%%%%%%%%%%%%  FORMALISM  %%%%%%%%%%%%%%%%%%%%%
%%%%%%%%%%%%%%%%%%%%%%%%%%%%%%%%%%%%%%%%%%%%%%%%%%%%%

%\setcounter{equation}{0}
\section{Kinetic theory of a hadronic gas mixture}

\subsection{Quantum Boltzmann equations}
We first explain our theoretical framework which is necessary for 
computing the shear viscosity coefficient. 
Consider a dilute gas of pions ($\pi$) and nucleons ($N$) in which
particles interact with each other through binary collisions. 
Nonequilibrium processes such as relaxation 
to thermal equilibrium can be described by kinetic equations
for one particle distribution functions 
$f^\pi ({\bf x, p}, t)$ and $f^N({\bf x, p}, t)$ (below, we suppress 
${\bf x}$-dependence). For simplicity, 
we assume that the gas is isospin symmetric, and thus 
$f^\pi ({\bf p}, t)$ and $f^N({\bf p}, t)$ are isospin averaged 
distributions. The relativistic 
quantum Boltzmann equations (more precisely, the Uehling-Uhlenbeck 
equations) of this hadronic gas mixture are then given by
\begin{eqnarray}
&&\hspace{-4mm}\frac{p^{\mu }}{E^{\pi }_{p}}\partial_{\mu}f^{\pi}({p}) 
= {\cal C}^{\pi \pi}\left[f^{\pi},f^{\pi} \right]
  + {\cal C}^{\pi N} \left[ f^{\pi},f^{N} \right]\, , \label{bol_pi}\\
&&\hspace{-4mm}\frac{p^{\mu }}{E^{N}_{p}}\partial_{\mu}f^{N}({p})
={\cal C}^{N N} \left[ f^{N},f^{N} \right] 
 +{\cal C}^{N \pi}\left[ f^{N},f^{\pi} \right] ,
\label{bol_N}
\end{eqnarray}
where 
$E^{\pi, N}_{p}=\sqrt{m_{\pi, N}^{2}+p^{2}}$ 
and the collision terms are defined as
\begin{widetext}
\begin{eqnarray}
 {\cal C}^{\pi \pi}+{\cal C}^{\pi N} 
&=&\frac{g_{\pi}}{2} \int d\Gamma^{\pi\pi}
 \Big\{   f^{\pi}_{1}f^{\pi}_{2} \left( 1+f^{\pi}_{3}\right) 
          \left( 1+f^{\pi}_{p}\right) 
       -
          \left( 1+f^{\pi}_{1}\right) \left( 1+f^{\pi}_{2}\right) 
          f^{\pi}_{3}f^{\pi}_{p}
 \Big\} 
\nonumber \\
&+& g_{N} \int d\Gamma^{\pi N}
 \Big\{   f^{N}_{1}f^{\pi}_{2} \left( 1-f^{N}_{3}\right) 
          \left( 1+f^{\pi}_{p}\right) 
       -
          \left( 1-f^{N}_{1}\right) \left( 1+f^{\pi}_{2}\right) 
          f^{N}_{3}f^{\pi}_{p} 
 \Big\}  \, , \label{col_pi}
\\
{\cal C}^{N N}+ {\cal C}^{N \pi}
&=&
 \frac{g_{N}}{2}\int d\Gamma^{NN}
 \Big\{   f^{N}_{1}f^{N}_{2} \left( 1-f^{N}_{3}\right) 
          \left( 1-f^{N}_{p}\right) 
        -
          \left( 1-f^{N}_{1}\right) \left( 1-f^{N}_{2}\right) 
          f^{N}_{3}f^{N}_{p}
 \Big\} 
\nonumber \\
&+&g_{ \pi } \int d\Gamma^{N \pi}
 \Big\{   f^{\pi}_{1}f^{N}_{2} \left( 1+f^{\pi}_{3}\right) 
          \left( 1-f^{N}_{p}\right) 
        -
          \left( 1+f^{\pi}_{1}\right) \left( 1-f^{N}_{2}\right) 
          f^{\pi}_{3}f^{N}_{p} 
 \Big\} \, .\label{col_N}
\end{eqnarray}
\end{widetext}
We have used shorthand notation 
$f^{\pi,N}_{i}\equiv f^{\pi,N}(k_{i}),\ f^{\pi,N}_{p}\equiv f^{\pi,N}(p)$
and $g_\pi,\ g_N$ are  the degeneracy factors $g_{\pi}=3$, $g_{N}=2$. 
For the collisions between the same species 
($\pi\pi$, $NN$), we have added a factor $1/2$. The factors 
$(1+f^\pi)$ and $(1-f^N)$ represent the Bose-Einstein and Fermi 
statistics of particles, respectively.
Finally, 
$d \Gamma$ in the integrants are invariant measures with the scattering 
amplitudes squared: For example, 
\begin{equation}
 d\Gamma^{\pi N}\equiv
|M_{\pi N}|^2 
\frac{(2\pi)^{4}\delta ^{(4)}(k_{1}+k_{2}-k_{3}-p) }
     {(2E^{N}_{1})(2E^{\pi}_{2})(2E^{N}_{3})(2E^{\pi}_{p})} 
\prod_{i}\frac{d^{3}k_{i}}{(2\pi)^3 } \, ,\label{gam}
\end{equation}
where $M_{\pi N}$ is the elastic scattering amplitude 
for $N(k_1)+\pi(k_2)\to N(k_3)+\pi(p)$. 
Explicit form of the scattering amplitudes will be shown at the end 
of this section. These are the basic ingredients of the kinetic theory.

Before we discuss how to solve the Boltzmann equations 
(\ref{bol_pi}) and (\ref{bol_N}), we need to know the 
equilibrium states. We define them without solving the full Boltzmann 
equations: They are given by the 
distributions $f^\pi_0,$ $f^N_0$ which make the collision terms 
vanish. Namely, 
${\cal C}^{\pi\pi}[f^\pi_0,f^\pi_0]
+{\cal C}^{\pi N}[f^\pi_0,f^N_0]={\cal C}^{N\pi}[f^N_0,f^\pi_0]
+{\cal C}^{N N}[f^N_0,f^N_0]=0$. 
These conditions are easily satisfied by the following 
Bose-Einstein and Fermi-Dirac distributions
if the common temperature $T=1/\beta$ and hydrodynamic 
velocity $V^\mu$ are used
\begin{eqnarray}
f^\pi_0(p)&=&\frac{1}{\e^{\beta V_\mu p^\mu }-1} \, ,\label{eq_pi}\\
f^N_0(p)&=&\frac{1}{\e^{\beta (V_\mu p^\mu-\mu)}+1}\, ,\label{eq_N}
\end{eqnarray} 
where $\mu$ is the baryon chemical potential.
The parameters $T$, $V^\mu$ and $\mu$ can, in principle,
 depend on the coordinates (local equilibrium), but when we compute 
quantities in thermal equilibrium, we simply select the rest frame 
$V^\mu=(1,0,0,0)$ so that $V_\mu p^\mu = E$.

In evaluating the entropy density, we use the expression in the 
equilibrium state as is done in the literature because the 
deviation from the equilibrium is assumed to be small. 
Namely, by using the grand partition functions for $\pi$ and $N$, 
\begin{eqnarray}
\ln Z_{\pi}=-Vg_{\pi} \int \frac{d^{3}p}{(2\pi)^{3}} 
\ln \Big(1-\e^{-\frac{E^{\pi}}{T}}\Big)\, , \\
\ln Z_{N}=Vg_{N} \int \frac{d^{3}p}{(2\pi)^{3}} 
\ln \Big(1+\e^{-\frac{E^{N}-\mu }{T}}\Big)\, ,
\end{eqnarray}
one obtains the total entropy density 
${ s}={ s}_{\pi}+{ s}_{N}$ as follows:
\begin{eqnarray}
{s}_{\pi}\!\!&=&\!\!
\frac{1}{V} \frac{\partial}{\partial T}T\ln Z_{\pi} \nonumber \\
\!\!&=&\!\!-g_{\pi} \int \frac{dp}{2 \pi^{2}} p^{2} 
\left\{ \ln \Big( 1-\e^{\frac{E^{\pi}}{T}} \Big) 
-\frac{E^{\pi}}{T(\e^{\frac{E^{\pi}}{T}}-1)} \right\}, \NN
{s}_{N}\!\!&=&\!\!\frac{1}{V} \frac{\partial  }{\partial T}T\ln Z_{N} 
\nonumber \\
\!\!&=&\!\!\! \! g_{N}\!\! \int\!\! \frac{dp}{2 \pi^{2}} p^{2} 
\left\{ \ln \Big( 1+\e^{\frac{E^{N}-\mu }{T}} \Big) 
+\frac{E^{N}-\mu }{T(\e^{\frac{E^{N}-\mu }{T}}+1)} \right\}.\nonumber
\end{eqnarray}

\subsection{Shear viscosity coefficient}
The shear viscosity 
coefficient is defined through the deviation of spatial components
of the energy momentum tensor in the linear order with respect to 
fluctuation from the equilibrium.
Consider a nonequilibrium state which is slightly deviated from 
the global equilibrium. 
Small deviation of the space components of energy 
momentum tensor $(T^{ij}=T^{ij}_0+\delta T^{ij})$ 
can be divided into traceful and traceless   parts:
\begin{eqnarray}
\delta T^{ij}&\equiv& \zeta\, (\delta ^{ij} \nabla \cdot  \mathbf{V})
   -2 \eta \left( \nabla ^{i} V^{j} \right)_{\rm trl}
           , \label{vdef}\\
\left( \nabla ^{i} V^{j} \right)_{\rm trl}
&\equiv &  \frac{1}{2}\Big(\nabla ^{i} V^{j}+\nabla ^{j} V^{i}\Big)
              - \frac{\delta ^{ij}}{3}\nabla \cdot  \mathbf{V} ,
   \label{traceless}
\end{eqnarray}
where $V^{i}$ is a space component of the hydrodynamic four velocity 
(which is common for $\pi$ and $N$)
$$
V^{\mu }=\frac{\int d^3p\, \frac{p^\mu}{E^{\pi,N}}\, f^{\pi,N}(p)}{\int d^3p f^{\pi,N}(p)}\, .
$$ 
Eq.~(\ref{vdef}) is the definition of the shear and bulk 
viscosity coefficients $\eta$ and $\zeta$. 
The flow vector $V^i$, as well as $T$ and $\mu$, is
in principle arbitrary and can depend on spatial coordinates.
But below we consider the case where only the flow vector $V^i$
depends on the coordinates, and in particular, its divergence 
is vanishing: $\nabla^i V^i = 0$ and $\nabla^i V^j\neq 0\, (i\neq j)$.
This situation corresponds to considering only the shear viscosity 
coefficient $\eta$ and ignoring all the other transport coefficients 
such as bulk viscosity or heat conductivity.

If one knows distributions 
$f^{\pi,N}(p)=f^{\pi,N}_0(p)+\delta f^{\pi,N}(p)$ as the solutions to 
the Boltzmann equations (\ref{bol_pi}) and (\ref{bol_N}), 
one can explicitly compute the (deviation of) energy momentum tensor:
\begin{eqnarray}
\delta T^{ij}
&=&g _{\pi} \int \frac{d^3p}{(2 \pi)^{3}}\frac{p^{i} p^{j}}{E^{\pi}_{p} } 
   \delta f^{\pi}{(p)}\NN
&+&g _{N} \int \frac{d^3p}{(2 \pi)^{3}}\frac{p^{i} p^{j}}{E^{N}_{p} } 
   \delta f^{N}{(p)}.
\label{ener}
\end{eqnarray} 
As mentioned above, we consider only the deviation 
$\delta f^{\pi,N}$ that originates from the shear. 
Then, it is quite convenient 
to parametrize $\delta f^{\pi}$ and $\delta f^{N}$ as follows 
($\hat p_i=p_i/p$):
\begin{eqnarray}
&&\delta f^{\pi}\equiv 
- f^{\pi}_{0}(1+ f^{\pi}_{0})\, \beta\, B^{\pi} (p) \NN
&&\qquad \qquad \times
 \left(\hat{p_{i}} \hat{p_{j}} -\frac{\delta _{ij}}{3} \right) 
 \left( \nabla ^{i} V^{j} \right)_{\rm trl} \, ,  \label{pide}   \\
&&\delta f^{N}\equiv
- f^{N}_{0}(1- f^{N}_{0})\, \beta\, B^{N} (p) \NN 
&&\qquad \qquad \times
 \left(\hat{p_{i}} \hat{p_{j}} -\frac{\delta _{ij}}{3} \right) 
 \left( \nabla ^{i} V^{j} \right)_{\rm trl} \, , \label{Nde} 
\end{eqnarray}
where we have introduced new quantities $B^{\pi,N}(p)$ to be determined by 
the Boltzmann equations.

Substituting Eqs.~(\ref{pide}) and (\ref{Nde})
into Eq.~(\ref{ener}) and comparing the result with Eq.~(\ref{vdef}),
one finds the shear viscosity coefficient $\eta$ as a function of 
unknown functions $B^{\pi,N}(p)$: 
\begin{eqnarray}
\eta &=&\frac{g_{\pi}\beta}{15} \int \frac{d^{3}p}{(2 \pi)^{3}} \frac{f^{\pi}_{0} (1+f^{\pi}_{0})}{E^{\pi}_{p}}\,  p^{2}  B^{\pi}(p)  \NN
&+&\, \frac{g_{N}\beta}{15} \int \frac{d^{3}p}{(2 \pi)^{3}} \frac{f^{N}_{0} (1-f^{N}_{0})}{E^{N}_{p}}\, p^2 B^{N}(p) \, . \label{shear}
\end{eqnarray}
The unknown functions $B^{\pi,N}(p)$ (or equivalently, the deviations 
$\delta f^{\pi,N}$) are numerically determined 
by solving the Boltzmann equations that are linearized with respect to 
$\delta f^{\pi,N}$. This procedure corresponds to the lowest order
 Chapman-Enskog method. 
Here we discuss only the outline of the procedure to solve the 
Boltzmann equations. More details are discussed in Appendix B. 

After the linearization, the Boltzmann equations (\ref{bol_pi}), 
(\ref{bol_N}) become a coupled linear equations for $B^\pi(p)$ and 
$B^N(p)$. Following Ref.~\cite{Dobado1}, 
we solve these equations in the functional space spanned 
by the orthogonal polynomials $\{ W_{(n)}(p),\ n=0,1,2,...\}$. 
Let us expand $B^{\pi,N}(p)$ by these bases:
\BQ
B^{\pi,N}(p)=\sum_{n=0}^\infty b_{(n)}^{\pi,N} W^{\pi,N}_{(n)}(p)\, ,
\label{expansion}
\EQ
where $W^{\pi,N}_{(n)}(p)$ is a polynomial of the order $n$,  and 
$b_{(n)}^{\pi,N}$ is the coefficient independent of $p$. 
Notice that $W^{\pi}_{(n)}(p)$ and $W^{N}_{(n)}(p)$ are not equivalent 
to each other. Indeed, they are defined so that they 
 satisfy the following different 
orthogonal conditions: 
\begin{equation}
\int\! \frac{d^{3}p}{(2 \pi)^{3}} 
\frac{f^{\pi}_0(1+f^{\pi}_0)}{E^{\pi}_{p}} 
\, p^{2}\, W^{\pi}_{(n)}(p) W^{\pi}_{(m)}(p)  = 
\delta_{nm} L^\pi_{(n)}  ,
 \label{ot1} 
\end{equation}
\begin{equation}
\int\! \frac{d^{3}p}{(2 \pi)^{3}} \frac{f^{N}_0(1-f^{N}_0)  }{E^{N}_{p}} 
\, p^{2}\,  W^{N}_{(n)}(p) W^{N}_{(m)}(p)  = 
\delta_{nm} L^N_{(n)} . \label{ot2}
\end{equation} 
The normalization factors $L^\pi_{(n)}$ and $L^N_{(n)}$ are not chosen 
to be 1. Instead, we choose the polynomial so that the coefficient 
of the term with the highest degree 
is unity (such polynomials are called "monic"). Namely, the first three 
polynomials have the following form:
\begin{eqnarray}
W_{(0)}(p)&=&1 \, ,\NN
W_{(1)}(p)&=&p+c_1 \, ,\NN
W_{(2)}(p)&=&p^2+d_1 p+d_2 \, .\nonumber
\end{eqnarray}   
Parameters $c_1$, $d_1$ and $d_2$ are uniquely determined 
 by the orthogonal conditions (thus independent of the dynamics). 
In fact, $W_{(n)}$ has $n$ unknown parameters, which are uniquely 
determined by $n$ orthogonal conditions with lower polynomials 
($W_{(m)},\ m=0,\cdots, n-1$).
Thus, $L^\pi_{(n)}$ and $L^N_{(n)}$ in Eqs.~(\ref{ot1}) and (\ref{ot2})
are known after we completely determine $W_{(n)}$.
Practically, the expansion (\ref{expansion}) is well approximated 
by the first few terms. Thus, in the present analysis, we take only 
the first three terms:
\begin{eqnarray}
&&\!\!\!\! B^{\pi}(p)\simeq  b^{\pi}_{(0)} + b^{\pi}_{(1)} W^{\pi}_{(1)}(p)
+ b^{\pi}_{(2)} W^{\pi}_{(2)}(p) \, ,\label{Bpi_exp}\\
&&\!\!\!\! B^{N}(p)\simeq b^{N}_{(0)}+b^{N}_{(1)}W^{N}_{(1)}(p)
+ b^{N}_{(2)}W^{N}_{(2)}(p)\, ,\label{BN_exp}
\end{eqnarray}
where we have used the definition $W_{(0)}^{\pi,N}=1$.
The coefficients $b_{(n)}^{\pi,N}$ are numerically determined. 
Once we know these coefficients, we can compute the shear viscosity 
coefficient $\eta$ from Eq.~(\ref{shear}).
We have checked that the results do not change even if we take 
up to the fourth terms $(n\le 3)$. 

Two comments are in order about our formulation. 
Firstly, we recall that the expansion of $B^{\pi,N}(p)$ in 
Eq.~(\ref{expansion}) is a familiar technique in solving the 
Boltzmann equation by the Chapman-Enskog method. 
If one treated a {\it classical} 
Boltzmann equation, measure of the orthogonal condition 
would be given by a much simpler distribution, the Maxwell-Boltzmann 
distribution. In this case, the polynomials that satisfy
the orthogonal condition are given by famous functions, 
the Sonine polynomials \cite{Chapman}. 
For the {\it quantum} Boltzmann equation, however, the measure
in the orthogonal condition is given by either the Fermi-Dirac 
or Bose-Einstein distribution, as shown in Eqs.~(\ref{ot1}) and
(\ref{ot2}). In this case, the polynomials 
satisfying them are not known, and we have to find them order 
by order.

Secondly, note that the shear viscosity coefficient $\eta$ defined by 
Eq.~(\ref{shear}) may be formally written as 
\BQ
\eta=\eta^\pi + \eta^N. \label{additive}
\EQ
It apparently looks `additive' with respect to each 
contribution. Indeed, if one considers a pion gas system without 
nucleons, one finds exactly the same expression as the first term of 
Eq.~(\ref{shear}). One might then be tempted to conclude from 
the expression (\ref{additive}) that the inclusion of nucleons
always contributes to enhance the value of shear viscosity. However, 
such argument does not make sense because the function $B^\pi(p)$
itself 
will change by the inclusion of nucleons. We will see later 
that $\eta^\pi$ indeed {\it decreases} as the effects of nucleons
become large (i.e., with increasing chemical potential).

\subsection{Scattering amplitudes of binary collisions}

Let us show the explicit expression of the scattering amplitudes
in the collision terms (\ref{col_pi}), (\ref{col_N}), and (\ref{gam}). 
Note that we have assumed factorization of the scattering 
amplitudes from the products of one-particle distributions 
in the collision terms. This is 
physically natural in a dilute gas where each collision is 
simply treated as an independent binary collision (ignoring 
higher order multiparticle correlations). Therefore, as long as 
we consider a dilute gas system where the Boltzmann equation is 
applicable, it is reasonable to use the amplitudes for two 
particle scatterings in the vacuum (free space). 
This is a great merit in computing transport coefficients. 
If one follows the microscopic Kubo formula to compute the transport 
coefficients, it is quite nontrivial how to include the 
effects of physical processes. On the other hand, if one uses 
the Boltzmann equation, one can easily incorporate the physical 
cross sections in the collision terms, which is however less justified from 
the first principle. 
In the present paper, we examine two different parametrizations 
for the scattering amplitudes. One is based on theoretical calculations, 
while the other is constructed from the experimental data. 
More precisely, we use the amplitudes from the low energy effective 
field theory (EFT) on the one hand, and the phenomenological 
amplitudes designed to reproduce experimental data of the 
elastic scatterings on the other hand. 

Scattering amplitudes based on EFT
are the following. First of all, the isospin-averaged
$\pi\pi$ scattering amplitude is 
given by the leading order Chiral Perturbation Theory (LO-ChPT)
\cite{Weinberg:1978kz}: 
\BQ\label{amp_pipi}
|M_{\pi\pi}|^2=\frac{1}{9f_\pi^4}\Big\{
21 m_\pi^4 + 9 {\sf s}^2 - 24 m_\pi^2 {\sf s} + 3({\sf t-u})^2
\Big\}\, ,
\EQ
where $f_\pi$ is the pion decay constant ($f_\pi=93$ MeV), 
${\sf s,\ t}$ and ${\sf u}$ are the Mandelstam variables for the scattering 
$\pi(k_1)+\pi(k_2)\to \pi(k_3)+\pi(p)$. 

Next, for the $\pi N$ scattering,
we use the results of low energy effective 
theory (LO heavy baryon ChPT) \cite{Fettes:1998ud}.
The isospin averaged $\pi N$ scattering amplitude in the 
Center-of-Mass (CM) frame is 
\begin{equation}\label{amp_piN}
|M_{\pi N}|^2=(2m_N)^2 \Big\{4 |g_-|^2 
+2 q_{\rm cm}^4 \sin^2 \theta_{\rm cm}\,
 |h_+|^2 \Big\}\, ,
\end{equation}
where $q_{\rm cm}$ and $\theta_{\rm cm}$ are, respectively, 
the magnitude of pion momentum, and the scattering angle in the CM frame. 
Two functions $g_-$ and $h_+$ are  
\BQA
&&g_-=-\frac{g_A^2}{f_\pi ^2}\frac{1}{4 \omega } 
\left(2\omega ^2-2m_{\pi}^2+ {\sf t} \right)+ \frac{\omega }{2f_{\pi}^2} 
\, ,\qquad \NN
&&h_+=-\frac{g_A^2}{f_{\pi}^2} \frac{1}{2\omega }\, ,\nonumber
\EQA
where $g_A=1.26$ is the nucleon axial charge,
 and $\omega$ is the pion energy in the CM frame.  
Note that the overall factor in Eq.~(\ref{amp_piN}) is not the same 
as in Ref.~\cite{Chen2} though both are based on 
Ref.~\cite{Fettes:1998ud}. This is because 
we have used the standard normalization factor for the 
spinors $\bar{u}^r (p) u^s(p)=2m_{N} \delta ^{rs}$ instead of 
$
\bar{u}^r(p) u^s(p)=\{(E^N + m_N)/{2m_N}\} \delta^{rs}
$
which was adopted in Ref.~\cite{Fettes:1998ud}.

Lastly, let us consider the $NN$ scattering. 
In the low energy EFT, scattering amplitudes in the CM frame 
for fixed spin and isospin can be parametrized in terms of the 
scattering length $a$ and the effective range $r$ as
\begin{eqnarray}\label{amp_NN_Swave}
\vert M_{NN'}\vert^2
=64\pi^2 {\sf s}\cdot 
\frac{1}{|-a^{-1}+\frac{1}{2}rq_{\rm cm}^2 - iq_{\rm cm}|^2}\, ,
\end{eqnarray}  
where $q_{\rm cm}$ is 
the magnitude of nucleon momentum in the CM frame. 
For each process, parameters $a$ and $r$ are determined to 
fit the low energy experimental data with the 
contribution from Coulomb force removed. Their numerical values 
are shown in table I. They are taken from Ref.~\cite{scef}:
%%%%%%%%%%%%%%%%%%%%%%%%%%%%% table %%%%%%%%%%%%%%%%%%%%%%%%%%%%%
\begin{center}
\begin{table}[h]
\caption{Scattering length $a$ and effective range $r$ for $NN$ collisions. }
\begin{tabular}{|c||c|r|r|} \hline
system & parameter & $S=0, I=1$ & $S=1, I=0$\\ \hline
$pp$& $a$& $-17.1$ (fm)& -----\hspace{6mm} \\
  & $r$& 2.79 (fm)& -----\hspace{6mm} \\
\hline
$nn$& $a$& $-16.6$ (fm)& -----\hspace{6mm} \\
  & $r$& 2.84 (fm)& -----\hspace{6mm} \\
\hline
$np$& $a$& $-23.7$ (fm)& 5.42 (fm)\\
  & $r$& 2.73 (fm)& 1.73 (fm)\\
\hline
\end{tabular} 
\end{table}
\end{center}
%%%%%%%%%%%%%%%%%%%%%%%%%%%%% table %%%%%%%%%%%%%%%%%%%%%%%%%%%%%

In low energy scattering, the dominant contribution to the
amplitude is given by the $s$-wave (orbital angular momentum $\ell=0$). 
Thus, we construct the spin-isospin averaged scattering amplitude 
from the expression (\ref{amp_NN_Swave}) for the $s$-wave 
with appropriate weight factors of spin and isospin. 
Details of the spin-isospin average are explained in Appendix A.

%%%%%%%%%%%%%%%%%%%%%%%%%%%%% figure %%%%%%%%%%%%%%%%%%%%%%%%%%%%%
\begin{figure}[t] 
     \includegraphics*[scale=1.0]{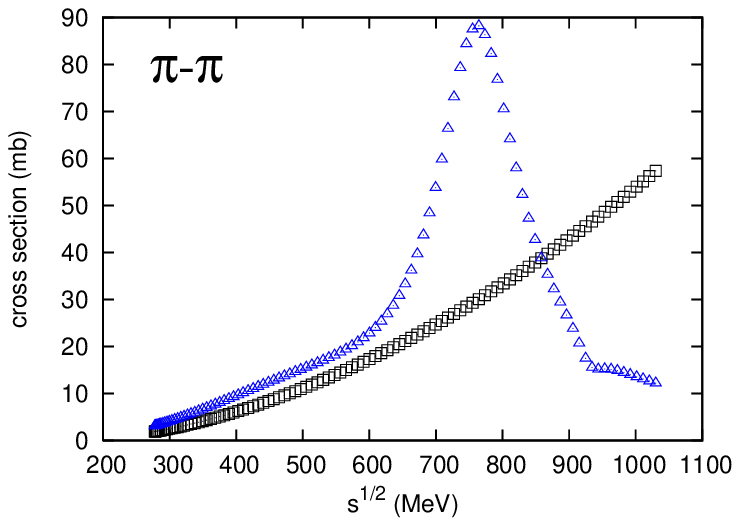}
     \includegraphics[scale=1.03]{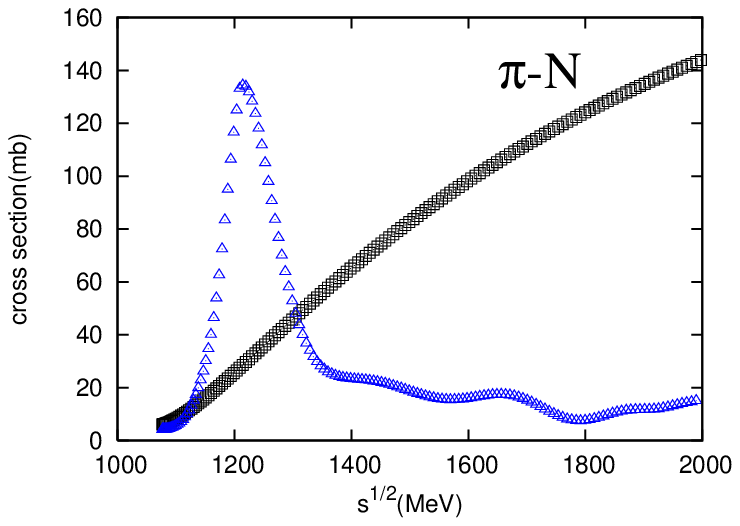}\\
\vspace{0mm}
\hspace{-4mm}
     \includegraphics*[scale=1.1]{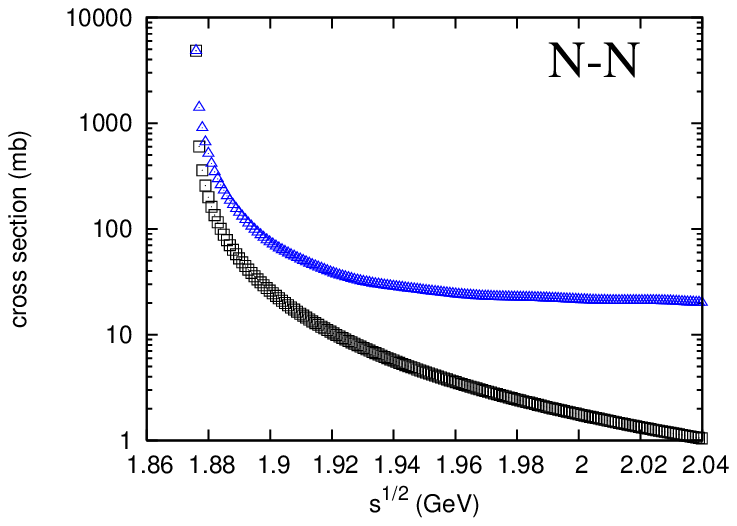}
     \caption{(Color online) Comparison of the elastic cross sections 
              in $\pi\pi$, $\pi N$, and $NN$ scatterings. Square 
              and triangle points are the results of the low energy EFT
               and the phenomenological amplitudes, respectively. 
               The actual fit to the experimental data was done for 
               the differential cross sections.
              The $NN$ elastic cross section is shown in logarithmic scale.}
     \label{Cross_secs}
\end{figure} 
%%%%%%%%%%%%%%%%%%%%%%%%%%%%% figure %%%%%%%%%%%%%%%%%%%%%%%%%%%%

In summary, the scattering amplitudes based on EFT are given by 
Eqs.~(\ref{amp_pipi}), (\ref{amp_piN}), and (\ref{amp_NN_Swave}).
It should be emphasized that all these expressions of the scattering 
amplitudes are valid only in a limited region of kinematics. 
LO-ChPT (for $\pi\pi$ case) is usually considered to be 
valid for $p\ll 4\pi f_\pi\sim$ 1 GeV with $p$ being the magnitude of 
pion momentum. In the heavy baryon ChPT for the 
$\pi N$ case, there is an additional expansion parameter 
$ p/m_N\ll 1$, which gives almost the same limitation 
as for the $\pi\pi$ case. 
Lastly, for the $NN$ case, the differential cross section
(\ref{amp_NN_Swave}) is valid only for small $q_{\rm cm}$, or 
equivalently, small scattering energy near the threshold. 
This fact implies that the validity of results derived from 
the Boltzmann equations will also be restricted to a small 
region in the $T$-$\mu$ plane, as we will discuss later.

 With this limitation in mind, it is quite important to 
consider physical scattering amplitudes in order 
to check the usefulness of, or more importantly to go beyond, 
the low energy EFT. To this end, 
we have constructed {\it phenomenological amplitudes} from 
the experimental data (differential cross sections) by fitting 
the coefficients of the partial-wave expansion. 
The fit was performed to the data up to 
$\sqrt{\sf s}=1.15$ GeV ($q_{\rm cm}=550$ MeV) in the $\pi\pi$
scattering, $\sqrt{\sf s}=2.00$ GeV ($q_{\rm cm}=770$ MeV) in the 
$\pi N$ scattering, and $\sqrt{\sf s}=2.04$ GeV ($q_{\rm cm}=405$ MeV)
in the $NN$ scattering.
Details of the fitting procedure are discussed in Appendix A. 
In Fig.~\ref{Cross_secs}, we compare elastic 
cross sections of the two different parametrizations. 
The elastic cross sections from the phenomenological amplitudes 
are almost identical with the experimental data (not shown) 
in the energy regions shown in the figures. Peaks in $\pi\pi$ and 
$\pi N$ cross sections are $\rho$-meson and $\Delta$ resonances, 
respectively. 
As mentioned above, the range of validity of LO-ChPT 
(for $\pi\pi,\, \pi N$ cases) is $p\ll $ 1 GeV for the pion momentum.
Thus, if one applies this limitation to the momentum of colliding 
particles in the CM frame, one finds for the scattering energy 
$\sqrt {\sf s} \ll \sqrt{\sf s_0}$ with $\sqrt{\sf s_0}\sim 2$ GeV 
for the $\pi\pi$ scattering, 
and $\sqrt{\sf s_0}\sim$ 2.4 GeV for the $\pi N$ scattering. 
Indeed, as evident from the figure, deviation of LO-ChPT 
from the phenomenological cross sections is already sizable 
well below the upper limits  
(mainly because of the $\rho$-meson and $\Delta$ resonances). 
For example, in the $\pi\pi$ case, LO-ChPT gives the same tendency 
as the phenomenological cross section up to 
$\sqrt{\sf s}\sim 600$~MeV ($q_{\rm cm}\sim 270$~MeV), 
but beyond that the deviation is not small. 
Therefore, it would be safe to consider the validity region of 
LO-ChPT to be 
\BQ\label{Validity_Limit_pipi}
\sqrt{\sf s}\ \simle\, \sqrt{{\sf s}_{\rm max}^{\rm EFT}}\equiv 600~{\rm MeV} 
\quad (\pi \pi\ {\rm scattering}),
\EQ
which is of course within 
$\sqrt {\sf s} \ll \sqrt{\sf s_0}\sim 2$ GeV.
Later we will use this limit to evaluate the maximum temperature 
up to which LO-ChPT is applicable.
 From the figures, it is obvious that the difference of 
two parametrizations becomes 
larger and larger with increasing energies. The $\pi\pi$ and 
$\pi N$ cross sections from LO-ChPT monotonically 
increase and become too large compared to the actual 
physical cross sections. We will see in the next section 
that this difference greatly affects the numerical value 
of the shear viscosity.

%%%%%%%%%%%%%%%%%%%%%%%%%%%%%%%%%%%%%%%%%%%%%%%%%%%%%
%%%%%%%%%%%%%%%%  NUMERICAL RESULTS  %%%%%%%%%%%%%%%%
%%%%%%%%%%%%%%%%%%%%%%%%%%%%%%%%%%%%%%%%%%%%%%%%%%%%%
\section{Numerical results} 
\subsection{Pion gas}
Let us first discuss the case with only pions. 
This is important for properly understanding the effects of 
nucleons in the next subsection and, 
at the same time, for checking the validity of our 
calculation compared with the existing results 
\cite{Chen1,Dobado3}. As we mentioned before, 
our formalism for the $\pi N$ gas can be easily reduced to 
the case with only pions (by setting the nucleon degeneracy factor 
$g_N \to 0$, for example). Then, the shear viscosity coefficient is 
given by the first term of Eq.~(\ref{shear}), and we determine 
$B^\pi(p)$ by solving the Boltzmann equation (\ref{bol_pi}) with 
the collision term given by the first term of Eq.~(\ref{col_pi}).
The expansion of $B^\pi(p)$ with respect to the orthogonal polynomials
was taken up to the third order 
to ensure the convergence of the result.

\subsubsection{Range of validity}

\begin{figure}[t]
\includegraphics*[scale=1.]{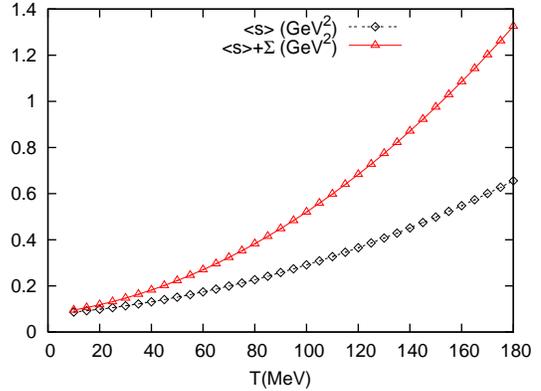}
\caption{(Color online) 
Temperature dependence of the average scattering 
energy squared $\langle {\sf s}\rangle$ and 
the measure of highest scattering energy squared 
${\sf s}_{\rm max}=\langle {\sf s}\rangle+ \Sigma$. }
\label{Average_energy}
\end{figure}

Before presenting our numerical results, we clarify the range
of validity in temperature for two parametrizations of the 
scattering amplitudes. 
As we already specified in Eq.~(\ref{Validity_Limit_pipi}), 
 LO-ChPT is valid only in a limited kinematical regime: 
$\sqrt{\sf s} \simle \sqrt{{\sf s}^{\rm EFT}_{\rm max}}=600$ MeV, 
while the phenomenological amplitude is by construction valid up to 
$\sqrt{\sf s}=1.15$~GeV. These conditions can be translated to 
the limitation in temperature in the following way.
Consider the binary collisions in thermal equilibrium. 
Since each collision takes place between particles in thermal 
equilibrium, the scattering energy squared ${\sf s}$ will fluctuate
around its mean value $\langle {\sf s}\rangle$ with a width 
$\Sigma$. The average $\langle {\sf s}\rangle$ and the width 
(standard deviation) $\Sigma$ may be defined by 
\BQA\label{average_s}
&&\hspace{-3mm}\langle {\sf s}\rangle \equiv 
\frac{\int \frac{d^3p_1}{(2\pi)^3}\int \frac{d^3p_2}{(2\pi)^3}\, 
{\sf s}(p_1,p_2)\, f^\pi_0(p_1)f^\pi_0(p_2)}
{\int \frac{d^3p_1}{(2\pi)^3}\int \frac{d^3p_2}{(2\pi)^3}\, 
f^\pi_0(p_1)f^\pi_0(p_2)}\, ,\\
&&\hspace{-3mm}\Sigma\equiv\sqrt{\langle {\sf s}^2 \rangle 
-\langle {\sf s}\rangle^2}\, .
\EQA
Because the scattering energy squared of most of the collisions are 
below ${\sf s}_{\rm max}(T)\equiv \langle {\sf s}\rangle  +\Sigma$,
one can regard ${\sf s}_{\rm max}(T)$ as (a measure of) the 
highest energy squared at temperature $T$.
So, one may interpret the validity condition of LO-ChPT 
as ${\sf s}_{\rm max}(T)< {\sf s}_{\rm max}^{\rm EFT}=0.36$ GeV$^2$, 
and thus obtain the validity condition for temperature 
$T<T_{\rm max}^{\rm EFT}$ with $T_{\rm max}^{\rm EFT}$ given by 
${\sf s}_{\rm max}(T_{\rm max}^{\rm EFT}) = 0.36$ GeV$^2$. 
In Fig.~\ref{Average_energy}, we show $\langle {\sf s}\rangle$ 
and ${\sf s}_{\rm max}$ as functions of temperature. 
 From the figure, one can read $T_{\rm max}^{\rm EFT}\sim 70$~MeV.
Therefore, we may conclude that the results of LO-ChPT are 
reliable only up to $T \sim 70$ MeV. On the other hand, 
the phenomenological amplitude is valid up to ${\sf s}\sim 1.3$ GeV$^2$. 
Therefore, we expect from the figure 
that the results of the phenomenological amplitude will be reliable 
up to temperature close to $T_c\sim 170$ MeV.

%%%%%%%%%%%%%%%%%%%%%%%%%%%%%%%%%%%%%%%%%%%%%%%%%%%%%%%
%%%%%%%%%%%%%%%%%%%%%%%%%%%%%%%%%%%%%%%%%%%%%%%%%%%%%%%

%%%%%%%%%%%%%%%%%%%%%%%%%%%%%%%%%%%%%%%%%%%%%%%%%%%%%%%
\begin{figure}[t]
\includegraphics*[scale=1.]{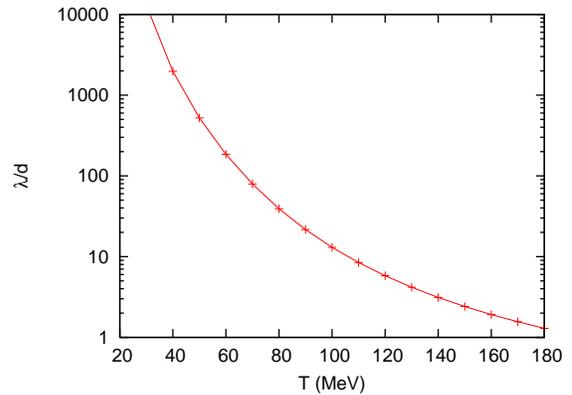}
\caption{(Color online) The ratio $\lambda/d$ as a function of 
temperature $T$. The description based on the Boltzmann equation 
is valid when $\lambda/d\gg 1$.
}
\label{Boltzmann_pi}
\end{figure} 
%%%%%%%%%%%%%%%%%%%%%%%%%%%%%%%%%%%%%%%%%%%%%%%%%%%%%%%

There is another important check for the validity of our framework. 
It is the applicability of the Boltzmann equations. 
Recall that in deriving the Boltzmann equations, one assumes
that a two-point correlation function can be factorized into 
a product of two one-point functions (one particle distributions 
$f(x,p,t)$). This is physically acceptable when the density of 
particles $n$ is small enough. This condition is normally expressed 
as
\BQ\label{Boltzmann_valid}
\lambda\gg d
\EQ 
where $\lambda$ is the mean-free path 
\BQ\label{mean-free-path}
\lambda=\frac{1}{n\sigma}
\EQ
with $\sigma$ being the cross section, 
and $d$ is the interaction range, meaning 
that each collision takes place independently. 
If we define the interaction range $d$ by the Compton length of pions: 
$d\sim 1/m_\pi$, then the validity condition for the Boltzmann 
equation reads  $\lambda/d \sim m_\pi/n\sigma\gg 1$. Alternatively, 
if we define $d$ %the interaction range  
through the cross section as
$\sigma\sim \pi d^2$, we obtain another expression
$\lambda/d\sim \sqrt{{\pi}/{\sigma^3}}/n \gg 1\, $. 
Both expressions give similar 
restriction on temperature. Figure \ref{Boltzmann_pi}
shows the ratio $\lambda/d$ as a function of temperature, 
where the (phenomenological) cross section $\sigma$ is 
estimated by its thermal average. One finds
 $\lambda/d\simge 3$ at $T=140$~MeV,
which manages to satisfy the inequalities. 
At this temperature,
the mean-free path of pions is estimated as $\lambda\sim 4$~fm, which is 
consistent with the literature \cite{MFPath}. As the temperature
is decreased, the mean-free path becomes longer, and the Boltzmann 
description gets better and better.

 Combining these two results, we may conclude that 
the phenomenological analysis of a pion gas based on the 
Boltzmann equation will be valid up to temperature $T\sim 140$~MeV 
which is much higher 
than the LO-ChPT limit $T\sim 70$~MeV, but lower than the 
critical temperature $T_c\sim 170$~MeV. 
However, we notice that there is a caveat to this conclusion. 
In fact, even though we have satisfactory description of $\pi\pi$ 
scatterings in wider kinematical regime, we have to worry 
about at least two other effects as temperature increases. 
The first one is the effects of other (heavier) degrees 
of freedom such as kaons, and the second is the possible
modification of pions in thermal enviornment. 
Both are however beyond the scope of the paper and we leave 
them for future problems.

\subsubsection{Shear viscosity coefficient $\eta$}

\begin{figure}[t]
\includegraphics*[scale=1.1]{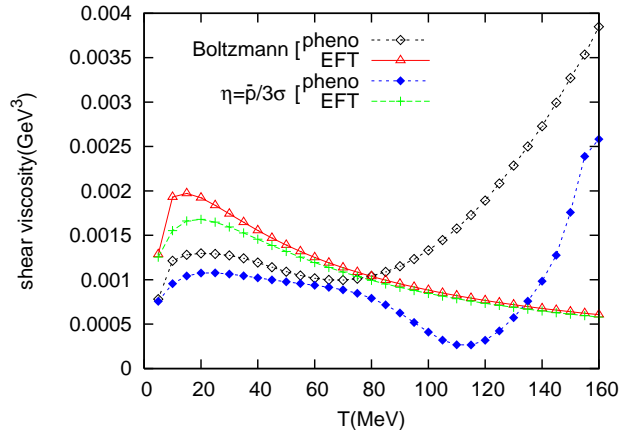}
\caption{(Color online) Shear viscosity coefficient $\eta$ of 
a pion gas as a function of temperature. Numerical results from 
the Boltzmann equations are compared with the rough estimate 
(\ref{rough_pipi}).}
\label{Viscosity_pipi}
\end{figure}

Figure \ref{Viscosity_pipi} shows the shear viscosity coefficient 
$\eta$ as a function of temperature $T$. 
Open diamonds ({\Large$ \diamond$}) and open triangles ($\triangle$)
are based on  numerical solutions to the Boltzmann equations 
with the phenomenological amplitudes and LO-ChPT, respectively. 
Remarkably, the two results show quite different behavior. 
While the result of LO-ChPT decreases with increasing $T$,
that of the phenomenological amplitude shows the opposite behavior. 

Qualitative behavior of two different results can be easily 
understood by using rough estimate of the viscosity coefficient. 
In classical transport theory for a dilute gas of one component, 
the shear viscosity coefficient is expressed in terms of the 
mean-free path $\lambda=1/n\sigma$, Eq.~(\ref{mean-free-path}), as follows:
\BQ\label{rough_pipi}
\eta \sim \frac13 n \overline p \lambda\, ,
\EQ
where $n$ is the particle number density, 
$\overline p$ is the average momentum, 
and $\sigma$ is the binary cross section. 
Thus, the shear viscosity coefficient is inversely proportional to 
the cross section. 
As we already saw in Fig.~\ref{Average_energy}, 
the typical energy in the $\pi\pi$ scattering increases 
with temperature. This means that average cross section 
$\langle \sigma \rangle$ indirectly depends on temperature. 
Notice also that the average momentum 
$\overline p=\int d^3p\, |{\bf p}|\, f_0^\pi(p)/\int d^3p f^\pi_0(p)$ 
roughly increases 
like $\overline p\propto \sqrt{T}$ % with increasing temperature 
(because $(\overline p)^2/2m\sim 3kT/2$). Therefore, temperature 
dependence of the shear viscosity coefficient is 
essentially determined by the interplay between those of 
$\overline p$ and $\langle \sigma \rangle$.  
For example, if the average cross section $\langle \sigma\rangle$ 
 increases rapidly as a function of $T$ 
(faster than $\sqrt T$), 
$\eta\sim \bar p/\langle\sigma\rangle$ is a decreasing function of $T$. 
But if $\langle \sigma \rangle$ is almost constant, 
$\eta$ is an increasing function of $T$. 
Based on these considerations and Fig.~\ref{Cross_secs}, 
we can easily deduce the followings:
As for the temperature dependence of the shear viscosity, 
we expect that $\eta_{\rm pheno}$ is slightly smaller than 
$\eta_{\rm ChPT}$ at low temperature 
(because $\sigma_{\rm pheno}\simge \sigma_{\rm ChPT}$), 
while $\eta_{\rm pheno}>\eta_{\rm ChPT}$  at high temperature. 
More precisely, since $\langle \sigma \rangle$ of 
the ChPT monotonically increases as $\langle \sigma\rangle 
\sim T^2/f_\pi^4$
while that of the phenomenological amplitude does not grow, 
we expect that the shear viscosity decreases in the ChPT case 
while the opposite happens in the phenomenological case.

%%%%%%%%%%%%% figure %%%%%%%%%%%%%%%%
%%%%%%%%%%%%%%%%%%%%%%%%%%%%%%%%%%%%%

Now let us come back to Fig.~\ref{Viscosity_pipi}, where we 
also show the rough estimate (\ref{rough_pipi}) with two 
different parametrizations. For the 
(average) cross section $\sigma$ in the rough estimate 
(\ref{rough_pipi}), we evaluate it at the average momentum 
$\overline p$, i.e.,  
$\langle \sigma \rangle \equiv 
\sigma(\overline p(T))$.\footnote{One can evaluate 
$\langle \sigma\rangle$ by its thermal average similarly 
as in Eq.~(\ref{average_s}), but the qualitative behavior is 
the same as $\sigma(\overline p(T))$.}
Comparing the results of the Boltzmann equations and of the 
rough estimate (\ref{rough_pipi}), we find that qualitative 
agreement of the results in two different ways of computation.
Nontrivial behavior of the rough estimate for the phenomenological 
amplitude (filled diamond points) is due to the resonance shape of the
elastic cross section in Fig.~\ref{Cross_secs}. Indeed, the valley 
around $T\sim 110$ MeV corresponds to the $\rho$ meson peak around 
${\sf s}\sim 0.5$ GeV$^2$ in the $\pi\pi$ cross section (According to 
Fig.~\ref{Average_energy}, ${\sf s}\sim 0.5$ GeV$^2$ is translated 
into $T\sim 150$ MeV, which is further diminished due to 
$\overline p\sim \sqrt{T}$ in the numerator of $\eta$). 
In the Boltzmann equations, such structure is further washed out 
by thermal average. From these analyses, we now understand that 
the behavior of the numerical results of the Boltzmann equations 
is largely due to the energy dependence of the cross sections
used in the collision terms. In particular, the decreasing $\eta$ 
of LO-ChPT at relatively high temperature $T\simge 80$~MeV is an 
artifact of too large cross section outside of the validity region 
of LO-ChPT. On the other hand, our most reliable result (open diamonds 
{\Large $\diamond$} in Fig.~\ref{Viscosity_pipi}) shows linear increase
with temperature for $T>120$~MeV. This behavior is consistent 
with that of hadronic resonance gas models \cite{Gyulassy,Gavin2}.
Moreover, our result is also consistent with that of 
Ultrarelativistic Quantum Molecular Dynamics (UrQMD) for 
a meson gas \cite{Muronga},
which also gives linear dependence on $T$, and 
$\eta\sim 0.1$~GeV$\cdot$fm$^{-2}=0.0039$~GeV$^{-3}$ 
at $T=150-160$~MeV.

\subsubsection{The ratio $\eta/s$}

\begin{figure}[t]
\includegraphics*[scale=1.1]{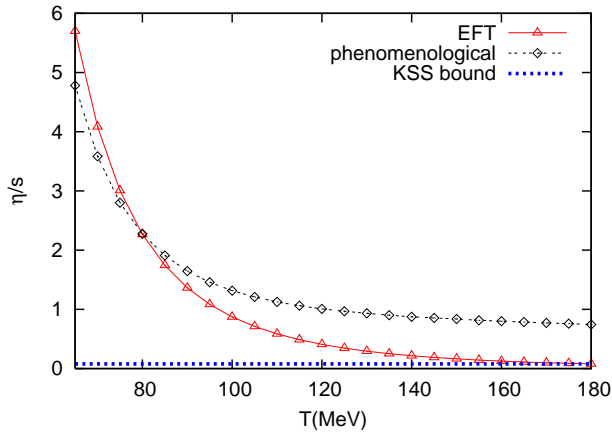}
\caption{(Color online) 
The ratio $\eta/s$ as a function of temperature $T$ 
in a pion gas system. Solid ($\triangle$) and dashed ({\Large $\diamond$}) 
curves are, respectively, the results with LO-ChPT and the 
phenomenological amplitude. Dotted line corresponds to the 
conjectured lower bound $\eta/s=1/4\pi$.}
\label{pionly}
\end{figure}

Figure \ref{pionly} shows our numerical results of the ratio 
$\eta/s$ as a function of 
temperature $T$. Dashed and solid lines correspond to the results of
LO-ChPT and the phenomenological amplitude, respectively. 
In both cases, $\eta/s$ is a monotonically decreasing function of $T$.
Notice that the result of LO-ChPT violates the conjectured 
bound $1/4\pi$ (the KSS bound, shown as the dotted line) at 
around the critical temperature $T_c\sim 170$ MeV. 
On the other hand, $\eta/s$ from the phenomenological amplitude
 keeps well above the KSS bound up to temperature $\sim T_c$. 
These are consistent with the results of 
Refs.~\cite{Chen1,Dobado3}.

Since the entropy density is common in both cases $s\propto T^3$, 
behavior of $\eta/s$ can be understood by that of $\eta$ itself. 
For example, we saw that $\eta$ decreases in LO-ChPT, while it 
increases in the phenomenological case. Such difference affects 
on the ratio $\eta/s$: it decreases faster in the 
LO-ChPT case than in the phenomenological one.
Also there is a crossing point for the two results of $\eta/s$ at 
$T\sim 80$~MeV (see Fig.~\ref{pionly}), and this point coincides 
with that of the shear viscosity $\eta$ (see the open diamonds 
and open triangles in Fig.~\ref{Viscosity_pipi}).

It has been argued in Ref.~\cite{Chen1} that 
violation of the KSS bound $\eta/s \ge 1/4\pi$ suggests the existence 
of phase transition (or crossover transition) in order for the KSS 
bound to remain valid. This kind of argument is of course dangerous 
because the precise value of $\eta/s$ depends on the amplitude in 
the collision term, and the result of Ref.~\cite{Chen1} 
is based on LO-ChPT, whose applicability is limited to 
$T \simle 70$~MeV as we already discussed in detail. In fact, 
more reliable result with the phenomenological amplitude does not 
violate the KSS bound even around the critical temperature. 
Therefore, ``violation of the KSS bound" cannot be the signature 
of phase transition.
 Instead of seeing violation of the KSS bound, 
we propose to check the other property to catch the indication 
of phase transition. As we mentioned 
in Introduction, Ref.~\cite{Kapusta} suggests that the ratio has 
a minimum around the (pseudo-) critical temperature $T_c$. 
If this is indeed the case, it implies that (for crossover transition) 
the curve of $\eta/s$ as a function of temperature will have a convex 
form around $T_c$, and the slope will decrease as temperature 
approaches $T_c$ from the left. (This kind of argument will hold
as far as the temperature is not too far away from the critical 
end point.) What we observed in Fig.~\ref{pionly} for the 
phenomenological amplitude is indeed the decrease of the slope 
with increasing temperature.  Thus, even if the ratio is still well
above the conjectured bound, we can anticipate the existence of phase 
transition.

%%%%%%%%%%%%%%%%%%%%%%%%%%%%%%%%%%%%%%%%%%%%%%%%%%
\begin{figure}[t]
\includegraphics*[scale=1.11]{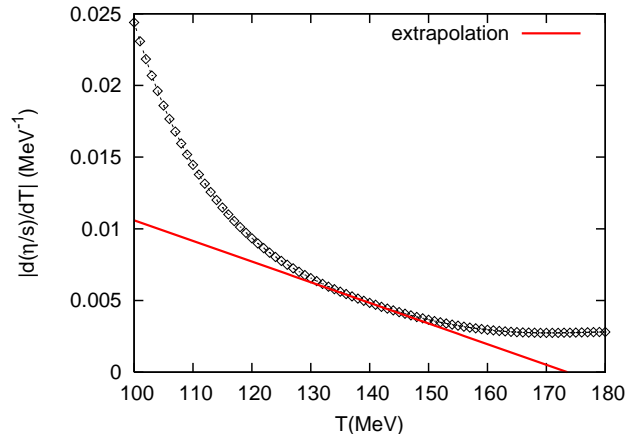}
\caption{(Color online) The absolute value of the slope $|d(\eta/s)/dT|$ 
for the result of phenomenological amplitude.  
Straight line is the linear extrapolation at $T=140$~MeV.
}
\label{slope}
\end{figure} 
%%%%%%%%%%%%%%%%%%%%%%%%%%%%%%%%%%%%%%%%%%%%%%%%%%

In Fig.~\ref{slope}, we show the (absolute value of) 
slope of the ratio $\eta/s$ for the phenomenological amplitude. 
Recall that we have estimated the border of validity region of our 
calculation to be $T\sim 140$~MeV.
Thus, we use the results around $T\sim 140$~MeV for linear
extrapolation towards higher temperature. More precisely, we
approximate the curve ${\cal R}(T)\equiv \eta/s$ around at some temperature 
$T=T_0<T_c$ as 
\BQ\label{expand}
{\cal R}(T)\simeq {\cal R}(T_0)
+{\cal R}'(T_0)(T-T_0)+\frac{{\cal R}''(T_0)}{2}(T-T_0)^2. 
\EQ
Then the slope $d{\cal R}(T)/dT$ is approximated 
by a linear function of $T$. The critical temperature $T_c$ may be 
 defined by the temperature where the slope is zero: 
$d{\cal R}(T)/dT=0$. Namely,
\BQ\label{expand1}
T_c \simeq T_0 - \frac{{\cal R}'(T_0)}{{\cal R}''(T_0)},
\EQ
where ${\cal R}'(T_0)<0$ and ${\cal R}''(T_0)>0$ on the left-hand side of 
a convex function ${\cal R}(T)$. The result of linear extrapolation
at $T=T_0=140$~MeV is shown on the same figure. 
Remarkably, the temperature at which the straight line cuts the 
horizontal axis is  173~MeV, which is quite a reasonable result 
as the critical temperature. In addition to this, 
one can guess the value of $\eta/s$ at $T = T_c$ 
by using the approximation (\ref{expand}):
\BQ\label{expand2}
{\cal R}(T_c)\simeq  {\cal R}(T_0) - \frac{({\cal R}'(T_0))^2}{2 {\cal R}''(T_0)}\, .
\EQ
If one substitutes numerical values at $T_0=140$~MeV, namely, 
${\cal R}(T_0)\simeq 1.0$, 
${\cal R}'(T_0)\simeq -0.005$~MeV$^{-1}$ and 
${\cal R}''(T_0)\simeq 1.38\times 10^{-4}$~MeV$^{-2}$  
which are read from Figs.~\ref{pionly} and \ref{slope}, 
one finds %the correction $\sim 0.1$ yielding 
${\cal R}=\eta/s\simeq 0.9$ at $T=T_c$. 
These results should be understood with reservation
at least for two reasons. First of all, there is ambiguity 
in the choice of $T_0$, which will affect the values of $T_c$ and 
 $\eta/s$. However, in fact, $T_0$ cannot be taken arbitrary 
because we need to take $T_0$ as close to $T_c$ as possible 
for the linear extrapolation to be accurate. To obtain reasonable 
values of $T_c$ and $\eta/s$, the largest possible value for $T_0$ 
is preferable. Since the choice $T_0=$140~MeV is the upper 
limit of our validity region, we expect that the estimated 
value $T_c=173$~MeV is the best value of our calculation. 
Therefore, even if there might be some ambiguity in selecting 
$T_0$, we can say that we have chosen the best value. 
The second source which might change the values of $T_c$ and 
$\eta/s$ is the possible contributions from heavier mesons.
Since such contributions become more important as $T \to T_c$,
our results with only pions become better as we depart from $T_c$
(which is however not desirable for determination of $T_c$).
Still, we expect our results are not so bad because at $T\sim 100$~MeV
such heavier particles can be ignored, and 
even at $T=140$~MeV, number of kaons 
amounts to only 20\% of total particles. 

For the crossover transition, the ratio $\eta/s$ will be continuous 
at the (pseudo) critical temperature $T_c$. This immediately implies
that the numerical value of $\eta/s$ determined above is relevant 
even in the deconfined phase. Our result $\eta/s\sim 0.9$ is well 
above the KSS bound, but is small enough compared to the weak-coupling 
QCD result (see for example, Fig.~4 of Ref.~\cite{Kapusta}). 
In this sense, the QCD matter around $T_c$ could be ``strongly 
interacting''. However, we should be careful when we draw such a
conclusion from the value of $\eta/s$. In fact, the smallness of 
the ratio $\eta/s$ is not a direct consequence of large cross section 
which may be realized by a strongly interacting matter.
In our calculation with the phenomenological amplitude, 
the cross section does not grow a lot (in contrast with the LO-ChPT) 
and the viscosity $\eta$ even {\it increases} as $T\to T_c$. 
Still, since the entropy increases faster than $\eta$, the ratio
$\eta/s$ becomes a decreasing function of temperature. 
Therefore, the smallness of the ratio is realized in a nontrivial way.

\subsection{Pion-nucleon gas}

Let us now present the numerical results for a dilute $\pi N$ gas. 
As advocated in Introduction, addition of nucleons to a pion gas
enables us to study the effects of baryon chemical potential. 
We discuss how the results of a pion gas presented in the previous 
subsection change under the influence of the chemical potential.

\subsubsection{Range of validity}

We start again by the discussion on the range of validity of 
our framework. Since we have to take the $\pi N$ and $N N$ 
collisions into  account in the $\pi N$ gas, we introduce the 
average scattering energy squared between particles $i$ and $j$ 
($i,j=\pi$ or $N$):
\BQ
\langle {\sf s}^{ij}\rangle \equiv 
\frac{\int \frac{dp_1^3}{(2\pi)^3}\int\frac{dp_2^3}{(2\pi)^3}
       \, {\sf s}^{ij}(p_1,p_2)\, f_0^i(p_1)f_0^j(p_2)}
     {\int \frac{dp_1^3}{(2\pi)^3}\int\frac{dp_2^3}{(2\pi)^3}
        f_0^i(p_1)f_0^j(p_2)}\, .
\EQ
For the $\pi\pi$ scattering, this is of course equivalent to 
Eq.~(\ref{average_s}) and depends only on $T$.
But for the $\pi N$ and $NN$ scatterings, the average values depend on
both $T$ and $\mu$. Similarly as before, we further define the 
standard deviation by 
$\Sigma^{ij} \equiv\sqrt{\langle ({\sf s}^{ij})^2 \rangle 
- \langle {\sf s}^{ij}\rangle^2}$, and regard 
${\sf s}_{\rm max}^{ij}(T,\mu)
\equiv \langle {\sf s}^{ij}\rangle + \Sigma^{ij}$ as a measure of 
the highest energy of the $i,j$ scattering at temperature $T$ and
baryon chemical potential $\mu$.

\begin{figure}[t]
\vspace*{-14mm}\hspace*{-10mm}
\includegraphics*[scale=0.85]{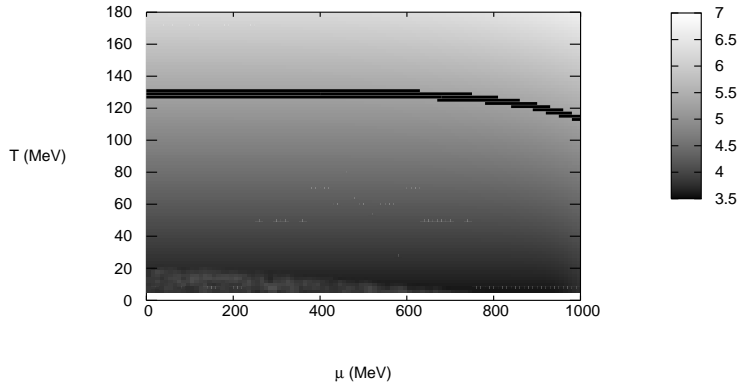}
\vspace{-12mm}
\caption{A measure of the highest scattering energy squared 
${\sf s}_{\rm max}^{NN}=\langle {\sf s}^{NN}\rangle+ \Sigma^{NN}$ for 
the $NN$ scattering is plotted in the $T$-$\mu$ plane. Gradation represents
the values in the range  3.5 GeV$^2<{\sf s}_{\rm max}^{NN}<\, $7 GeV$^2$. 
Solid curve corresponds to the borderline 
${\sf s}_{\rm max}^{NN}=5.29$~GeV$^2$.}
\label{Average_s_piN}
\end{figure}

In Fig.~\ref{Average_s_piN}, we show the values of 
${\sf s}^{NN}_{\rm max}$ 
on the $T$-$\mu$ plane. Recall that the fit to experimental data 
in the $NN$ scatterings is by construction valid up to 
${\sf s}^{NN}<(2.04)^2=4.16$~GeV$^2$. However, in fact, 
our parametrization works well up to slightly higher 
value ${\sf s}^{NN}\sim (2.3)^2=5.29$~GeV$^2$.
Therefore, we define the borderline of the validity region by 
temperature and chemical potential that satisfy 
${\sf s}_{\rm max}^{NN}(T,\mu)=5.29$~GeV$^2$. The result is shown 
on the same figure~\ref{Average_s_piN} by a thick solid curve. 
The maximum temperature is about 130~MeV or slightly smaller than that,
while the chemical potential is not restricted up to $\mu=1$~GeV. 
If we perform the same analysis for the $\pi N$ scattering, the 
borderline defined by ${\sf s}^{\pi N}_{\rm max}=4$~GeV$^2$ locates
outside of the region of our interest $T<180$~MeV, $\mu<1$~GeV.
On the other hand, the validity region of the low energy EFT is 
severely restricted. For example, if we take the maximum 
value of $\sqrt{{\sf s}^{NN}}$ to be 1.88~GeV 
(beyond which the phenomenological fit starts to deviate 
from the low energy EFT fit, see Appendix A), 
then the borderline defined by 
${\sf s}_{\rm max}^{NN}=(1.88)^2=3.53$~GeV$^2$ is very close to 
the horizontal axis (see Fig.~\ref{Average_s_piN}). 
Even if we relax the condition to higher value 
${\sf s}^{NN}_{\rm max}=(1.90)^2=3.61$~GeV$^2$, 
allowed region is still very narrow.

%%%%%%%%%%%%% figure %%%%%%%%%%%%%%%%
\begin{figure}[t]
\begin{center}
\includegraphics[scale=1.1]{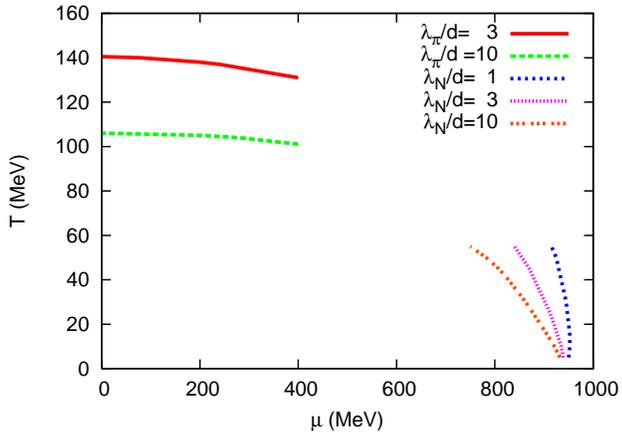}
\caption{
(Color online) The lines for the ratio 
$\lambda_{\pi,N}/d$ in a $\pi N$ gas. 
In the region of low density and high temperature, we show only 
$\lambda_\pi/d$, while in the region of low temperature and high 
density, $\lambda_N/d$. We may regard the line for $\lambda/d=3$
as the border of the validity region of the Boltzmann equations.
The line for $\lambda_\pi/d=1$ is far outside of the $T$-$\mu$ region
shown here. 
}
\label{Validity_Bol_piN}
\end{center}
\end{figure} 
%%%%%%%%%%%%%%%%%%%%%%%%%%%%%%%%%%%%%

Let us also examine the validity condition for the Boltzmann equations. 
The criterion is again given by Eq.~(\ref{Boltzmann_valid}), but now 
pion's mean-free path should be modified in the presence of nucleons,
and we have to separately consider nucleon's mean-free path, too. 
According to classical transport theories, 
the mean-free path of $i$-th component in a mixed gas is 
modified as \cite{Kennard} 
\BQ\label{modMFP}
\lambda_i= \frac{\lambda_i^0}{1+\sum_{j \ne i}
         \sqrt{\frac{1+m_i/m_j}{2}}
         \frac{\sigma _{ij}}{\sigma _{ii}}
         \frac{n_j}{n_i}}\, ,
\EQ
where $\lambda_i^0=1/n_i\sigma_i$ is the mean-free path of a pure gas 
of $i$-th component, and $\sigma_{ij}$ is the cross section between 
$i$-th and $j$-th components. Therefore, the mean-free paths of pions 
and nucleons are respectively given by 
\BQA
&&\lambda_\pi=\frac{1}{n_\pi \sigma_{\pi\pi}
                       +n_N\sigma_{\pi N}\sqrt{\frac{1+m_\pi/m_N}{2}}}\, ,
 \label{modMFP_pi}\\
&&\lambda_N=\frac{1}{n_N \sigma_{NN}
                      + n_\pi \sigma_{\pi N}\sqrt{\frac{1+m_N/m_\pi}{2}}}\, .
 \label{modMFP_N}
\EQA

Consider the condition for nucleons.
If one takes $\sigma_{NN}\sim 40$~mb as a typical value of the 
$NN$ cross section (twice of the saturating value at high energy, 
see Fig.~\ref{Cross_secs}) and uses $d\sim 1/m_\pi$ again 
for the interaction range, then the condition for nucleons 
$\lambda_N\gg d$ can be estimated as 
$n_N\ll m_\pi/\sigma_{NN}\simeq n_N^0$ with
$n_N^0=0.157$~fm$^{-3}$ being the normal nuclear density. 
This condition is physically quite 
reasonable since we do not expect standard Boltzmann description useful
at normal nuclear matter density. To obtain more precise restriction
depending on $T$ and $\mu$, we need to use Eqs.~(\ref{modMFP_pi}) and 
(\ref{modMFP_N}) and estimate each condition by replacing 
$\sigma$ and $n$ by their thermal averages. 
In Fig.~\ref{Validity_Bol_piN}, we show the lines for 
several values of the ratio $\lambda/d$ in the $T$-$\mu$ plane. 
For simplicity, we used $d=1/m_\pi$ for both pions and nucleons. 
In the region of low baryon density where pions are dominant degrees 
of freedom for transport phenomena, 
we show the ratio for pions $\lambda_\pi/d$. 
The line for $\lambda_\pi/d=3$ (solid line) is consistent 
with the previous result shown in Fig.~\ref{Boltzmann_pi}.
We may regard this line as the border of the validity of the Boltzmann 
equations at low densities. On the other hand, at large chemical 
potential and low temperature, transport phenomena is dominated 
by nucleons. The lines for the ratio $\lambda_N/d=1,\, 3,\, 10$ 
are shown in the figure. We may again regard the line $\lambda_N/d=3$ 
as the borderline, which reaches at $(T,\mu)=(0,\, 950{\rm MeV})$. 
This condition is more restrictive than that of Fig.~\ref{Average_s_piN}.

Combining these two analyses, we may conclude that our 
framework is valid in a wide region of the $T$-$\mu$ plane, whose 
boundary is roughly given by (a quarter of) the elliptic curve 
connecting $(T,\mu)\sim$~(130{\rm MeV},\, 0) and (0,\, 950{\rm MeV}).
The first point (130{\rm MeV},\, 0) is specified by the measure of the 
highest scattering energy squared, and the last point (0,\, 950{\rm MeV})
is from the validity limit of the Boltzmann equations.

%%%%%%%%%%%%% figure %%%%%%%%%%%%%%%%
\begin{figure}[t]
\begin{center}
\includegraphics[scale=1.1]{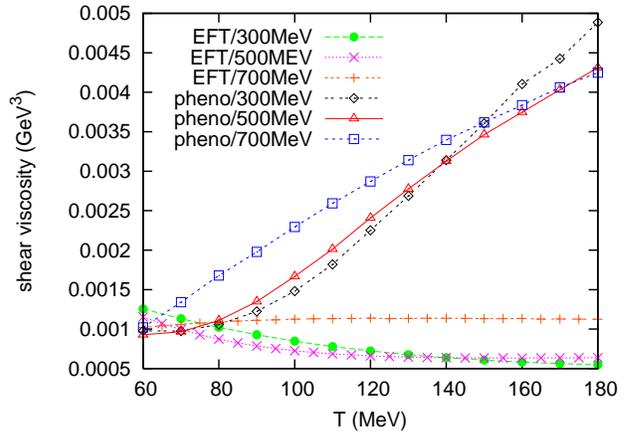}
\caption{
(Color online) 
Temperature dependence of the shear viscosity 
coefficient $\eta$ of a $\pi N$ gas at different values of 
chemical potential $\mu=300,\, 500,\, 700$ MeV. Comparison 
is made between the results of the low energy EFT and the 
phenomenological amplitudes.}
\label{Viscosity_piN}
\end{center}
\end{figure} 
%%%%%%%%%%%%%%%%%%%%%%%%%%%%%%%%%%%%%

%%%%%%%%%%%%% figure %%%%%%%%%%%%%%%%
\begin{figure*}[t]
\begin{center}
{\Large \hspace*{10mm}T~=~50~MeV   \hspace{5.5cm} T~=~100~MeV}\\
\includegraphics[scale=1.1]{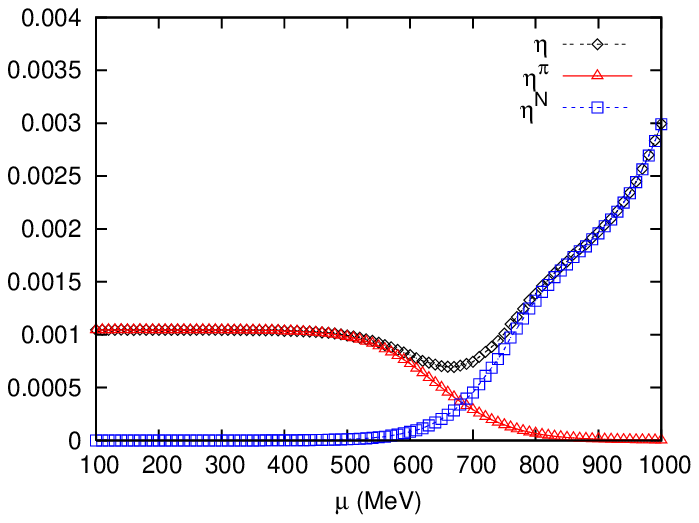}
\includegraphics[scale=1.1]{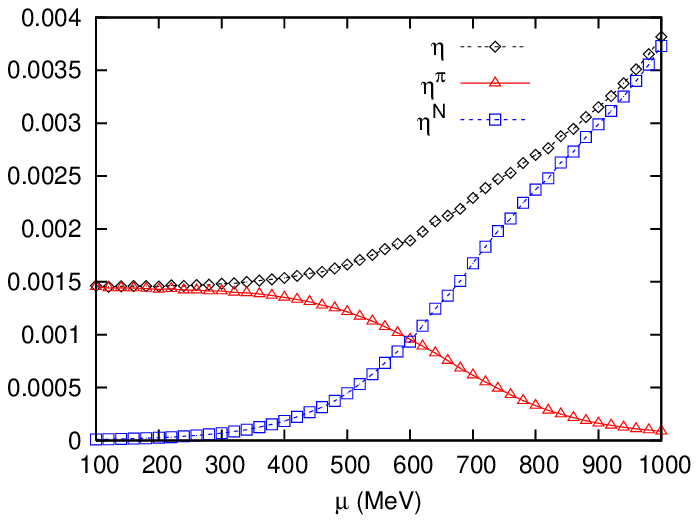}
\caption{(Color online)
$\mu$ dependence of $\eta$ at $T=50$MeV (left) 
and $T=100$MeV (right), and its decomposition 
$\eta=\eta^\pi + \eta^N$.}
\label{Viscosity_mu_dep}
\end{center}
\end{figure*} 
%%%%%%%%%%%%%%%%%%%%%%%%%%%%%%%%%%%%%

\subsubsection{Shear viscosity coefficient $\eta$}

In Fig.~\ref{Viscosity_piN}, we show temperature dependence 
of $\eta$ at different values of chemical potential 
$\mu=300,\, 500,\, 700$ MeV. Comparison is made between the 
results of the low energy EFT and the phenomenological amplitudes. 
These temperature dependence is qualitatively consistent with 
the previous results of the pion gas (see Fig.~\ref{Viscosity_pipi}). 
As for the $\mu$ dependence, however, the shear viscosity coefficient 
shows nontrivial behavior. In particular, in the window 
80~MeV $<T<$130~MeV, it {\it increases} with increasing $\mu$. 
This is not quite understandable at first 
because we expect that the cross 
section of $\pi\pi$ scattering will effectively enhance in 
the presence of nucleons while the effects of nucleon viscosity 
may be ignored at lower density, meaning that the shear 
viscosity will decrease. To understand what really happens
when $\mu\neq 0$, let us again consider a rough estimate of the 
shear viscosity of a mixed gas. Let the shear viscosity coefficient 
of a pure gas of particle species $i$ be 
$\eta^i_0=n_i \overline p_i \lambda_i^0/3$ where $n_i$ is the number 
density, $\overline p_i$ is the average momentum, and $\lambda_i^0$
is the mean-free path. Then the shear viscosity coefficient for 
$n$ component gas is given by the sum of each viscosity $\eta^i_0$
with modified mean-free path $\lambda_i$ given in Eq.~(\ref{modMFP})
\cite{Kennard}:
$$
\eta_{\rm mix}
=\sum_{i} \eta^{i}_0 \frac{\lambda_{i}}{ \lambda_{i}^0}\, .
$$
Thus for the $\pi N$ gas mixture, 
we obtain
\BQA\label{rough_piN}
\eta &=&\eta^\pi+\eta^N\\
&\simeq&\!\! \frac{\eta^{\pi}_0}
      {1+\frac{1}{\sqrt2}
         \left(\frac{\sigma_{\pi N}}{\sigma_{\pi\pi}}\right)
         \left(\frac{n_N}{n_\pi}\right)}
+\frac{\eta^N_0}
      {1+\sqrt{\frac{m_N}{2m_\pi}}
         \left(\frac{\sigma_{\pi N}}{\sigma_{NN}}\right)
         \left(\frac{n_\pi}{n_N}\right)}\, ,\nonumber
\EQA
where 
we have used the approximation $m_\pi/m_N\ll 1$.
In the two extreme limits $n_N/n_\pi\to 0$ and $\infty$, the formula 
(\ref{rough_piN}) reduces to $\eta^\pi_0$ and $\eta^N_0$, respectively.
Therefore, this formula interpolates a pure pion gas at low $\mu$ 
and a pure nucleon gas at high $\mu$.
Notice that the pion contribution $\eta^\pi$ is always smaller than 
$\eta^\pi_0$ due to the presence of $\pi N$ interaction, as 
we alluded before in relation to Eq.~(\ref{additive}). 
This is exactly what we expected. On the other hand, it is not 
straightforward to predict the behavior of the total shear viscosity.
If one plots the rough estimate (\ref{rough_piN}) as a function 
of $n_N/n_\pi$ assuming that the cross sections are constant 
and are of the same order, one finds that $\eta$ decreases at 
small values of $n_N/n_\pi$ (small $\mu$), but turns into 
increase at large $n_N/n_\pi$ (large $\mu$). If one changes the 
numerical value of cross sections a little, then the curve turns
into monotonic increase. In fact, both can happen in reality 
depending on temperature. In Fig.~\ref{Viscosity_mu_dep}, 
we have plotted $\eta$ at $T=50$~MeV and 100~MeV as 
a function of $\mu$. 
While the total viscosity behaves differently in these two panels, 
one can see monotonic decrease (increase) of $\eta^\pi$ ($\eta^N$), 
namely, the interplay between $\eta^\pi$ and $\eta^N$ at both temperatures.
Therefore, the increase of $\eta$ with increasing $\mu$ 
observed in Fig.~\ref{Viscosity_piN} can be understood as
a result of enhancement of $\eta^N$.

\subsubsection{The ratio $\eta/s$}

%%%%%%%%%%%%% figure %%%%%%%%%%%%%%%%
\begin{figure*}[t]
\begin{center}
\hspace*{-8mm}
\includegraphics[height=6.4cm,width=8.3cm]{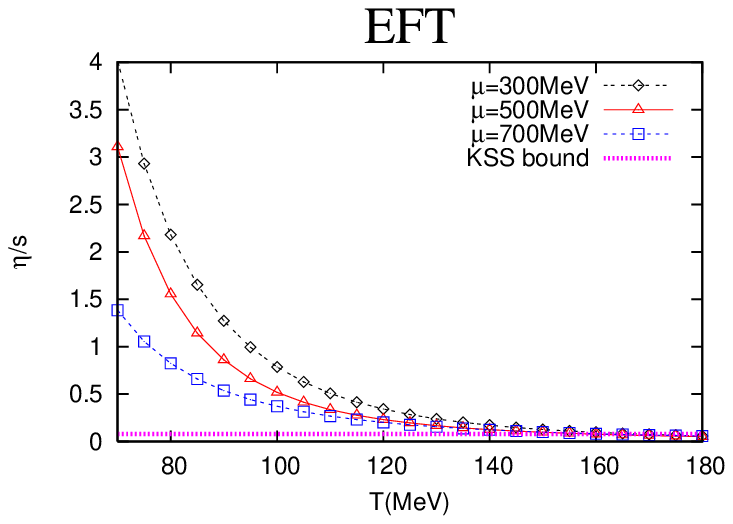}
\hspace{-0mm}
%\vspace{2.3cm}
\includegraphics[height=6.4cm,width=8.3cm]{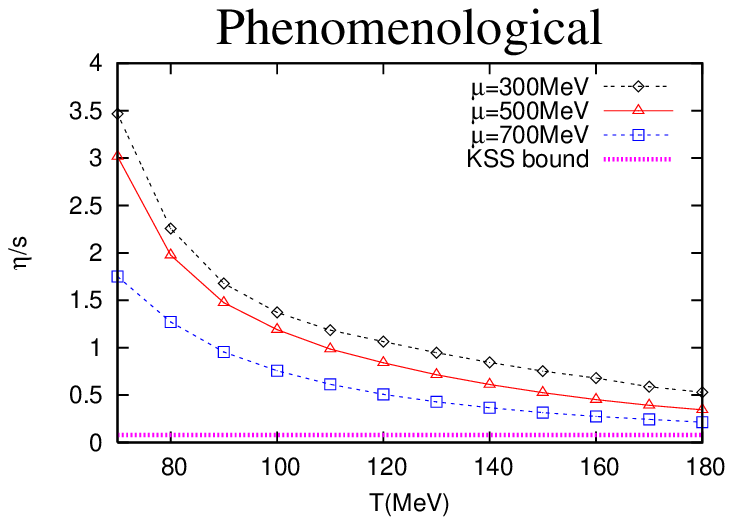}
\hspace{-4mm}
\caption{(Color online) 
The ratio $\eta/s$ of a $\pi N$ system plotted as a function 
of temperature $T$ at different values of baryon chemical potential
$\mu=300,\, 500,\, 700$ MeV. 
(Left) results of low energy EFT, 
(Right) results of phenomenological amplitudes. } 
\label{eta/s_Tdep}
\vspace{-0.6cm}
\end{center}
\end{figure*}
%%%%%%%%%%%%%%%%%%%%%%%%%%%%%%%%%%%%

Figure \ref{eta/s_Tdep} shows temperature dependence of the 
ratio $\eta/s$ of the $\pi N$ gas at different values of baryon
chemical potential $\mu=300,\, 500,\, 700$ MeV. 
The left-hand side is the results of the
low energy EFT, while the right-hand side, the phenomenological 
amplitudes.  
The ratio $\eta/s$ shows qualitatively the same behavior as 
in the pion gas system (Fig.~\ref{pionly}): (i) Both the results 
(EFT and phenomenological amplitudes)
are monotonically decreasing functions of $T$, (ii)
$(\eta/s)_{\rm EFT}>(\eta/s)_{\rm pheno}$ at lower $T$ while
 opposite at higher $T$, and (iii) the ratio of the EFT violates the 
KSS bound at around $T\sim T_c$ while that of the 
phenomenological amplitudes does not. Notice that 
the inclusion of chemical potential works to 
reduce the value of $\eta/s$. As a result, for the ratio of 
the EFT, the temperature at which the curve cuts the KSS
bound becomes smaller. On the other hand, for the result of the 
phenomenological amplitudes, the flattening of the curves seems 
to occur at lower temperature with increasing $\mu$. If this is 
indeed the symptom of crossover transition as we discussed before, 
one can say that the (pseudo) critical temperature $T_c$ will 
decrease with increasing $\mu$, which is consistent with what we know
from lattice simulations or effective models.~\footnote{One can try 
to determine the values of $T_c$ from the slope of each curve as 
in the pion gas case, but 
we do not perform such extrapolation for two reasons: First, 
it is practically very difficult to obtain reliable results 
since the highest temperature allowed for the Boltzmann 
equations gradually decreases with increasing $\mu$, as shown 
in Fig.~\ref{Validity_Bol_piN}. 
Second, the chiral phase transition at moderate values of $\mu$ 
will be most likely of the first order, for which  the ratio will show a 
discontinuity and the slope will not necessarily zero at $T_c$. 
Position of the critical end point may be detected from the 
behavior of the ratio $\eta/s$ \cite{endpoint}, but we do not 
expect our Boltzmann equations can 
describe the precise structure of the phase transition. 
We will discuss similar problems in relation to the nuclear 
liquid-gas phase transition in Sect.III.B.4. }

Our result $(\eta/s)_{\rm pheno}\sim 0.5-0.4$ at 
$\mu=700{\rm MeV}$ and $T\sim 100-140$ MeV is consistent with 
that of ``URASiMA" (Ultra-Relativistic AA collision Simulator 
based on Multiple scattering Algorithm)\cite{Muroya} 
which is a Monte-Carlo event generator of hadronic collisions. 
It includes both elastic and 
inelastic scatterings whose cross sections are given by experimental 
data. The numerical coincidence of the ratio from different 
frameworks is very interesting. In fact, the essential difference 
of URASiMA from our framework is the presence of inelastic collisions.
But as far as we consider small deviation from thermal 
equilibrium, inelastic collisions which will change particle 
numbers (such as $\pi\pi\to \pi\pi\pi$ or $\pi N\to \pi\pi N$) 
would not be so important, and it seems reasonable to obtain the 
same result from two different frameworks.

%%%%%%%%%%%%%%%%% figure %%%%%%%%%%%%%%%%%%%%
\begin{figure*}[t]
\begin{center}
\hspace*{-7cm}{\Large Phenomenological}\\
\hspace*{-6mm}\vspace{-10mm}\\
\includegraphics[height=5.8cm,width=8.3cm]{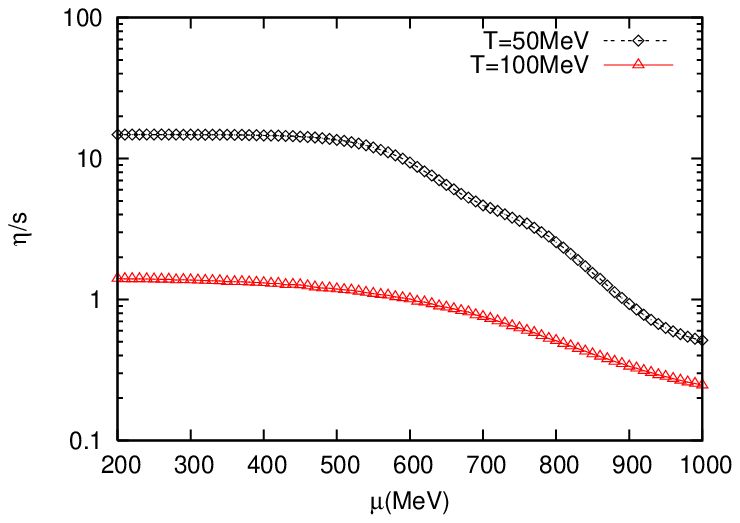}
\hspace{-4mm}
%\vspace{2.3cm}
\includegraphics[height=6.4cm,width=8.3cm]{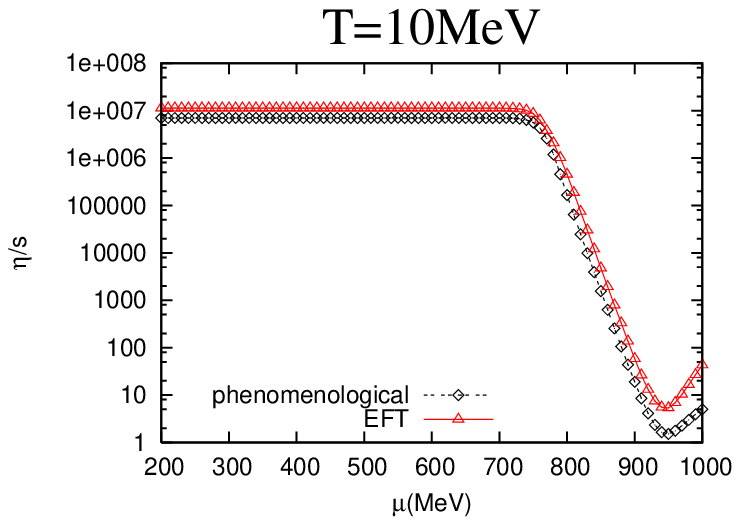}
\hspace{-4mm}
\caption{(Color online) $\mu$ dependence of $\eta/s$:
(left): the results of the phenomenological amplitudes 
at higher temperature $T=50,~100$~MeV,  
(right): the results at lower temperature $T=10$~MeV.} 
\label{eta/s_mu_dep}
%\vspace{-0.6cm}
\end{center}
\end{figure*}
%%%%%%%%%%%%%%%%%%%%%%%%%%%%%%%%%%%%%%%%%%%%%

As we already saw in Fig.~\ref{Viscosity_piN}, the shear viscosity 
coefficient {\it increases} in the window 80~MeV$<T<$130~MeV with 
increasing $\mu$. However, in the same window, the ratio {\it does}
decrease. This clearly implies that the reduction of $\eta/s$ 
at $T\sim 80-130$ MeV is due to the increase of entropy. 

In Fig.~\ref{eta/s_mu_dep}, we show the $\mu$ dependence of the 
ratio $\eta/s$ at temperature $T=50, 100$~MeV (left) 
and at low temperature $T=10$~MeV (right). 
Recall that the ratio decreases with increasing $\mu$ in 
Fig.~\ref{eta/s_Tdep} where $T\ge 70$~MeV was shown. This is 
consistent with the left figure, and in agreement with our 
original expectation as we mentioned in Introduction. 
On the other hand, the right figure shows a new structure: 
There is a valley at large $\mu\sim 950$~MeV. We will discuss 
later the physical implication of this structure.

Let us comment again the point made in the last paragraph of 
the previous subsection. The ratio $\eta/s$ becomes less than 0.3 
at $T\simge 140$~MeV and $\mu=700$~MeV. This is small enough and is 
close to the KSS bound $\eta/s\sim 0.1$ compared to other systems 
such as water. However, as shown in Figs.~\ref{Viscosity_piN}
and \ref{Viscosity_mu_dep}, the shear viscosity itself grows as 
the system approaches phase boundary ($T\to T_c,\, \mu\to \mu_c$). 
Therefore, even if the ratio is small enough, the $\pi N$ system 
cannot be treated, strictly speaking, as a perfect fluid with $\eta=0$. 
Sometimes ideal hydrodynamics 
is used to describe the matter after hadronization in heavy-ion 
collisions, but  one will have to take into account the effects of 
viscosity for a realistic simulation.

\subsubsection{Valley structure at large $\mu$ and low $T$}

\begin{figure*}[t]
\begin{center}
\hspace*{2cm}
{\Large EFT\hspace{6cm}Phenomenological}\\
\hspace*{-6mm}
\includegraphics[height=6.1cm]{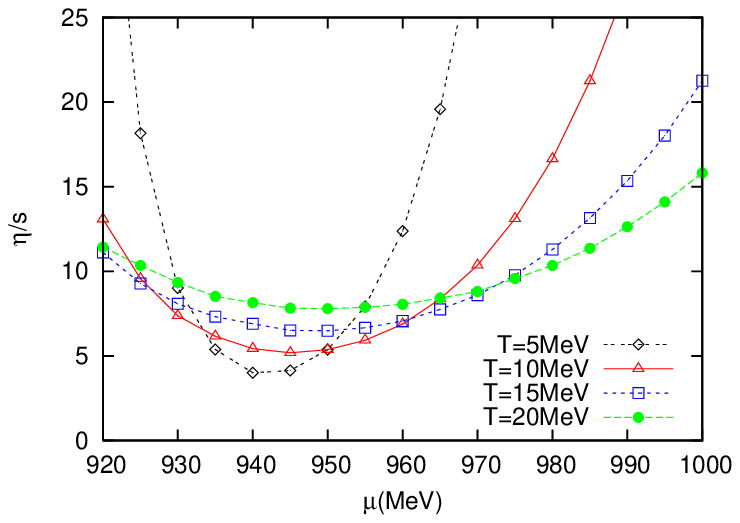}
\hspace{-4mm}
\includegraphics[height=6.1cm]{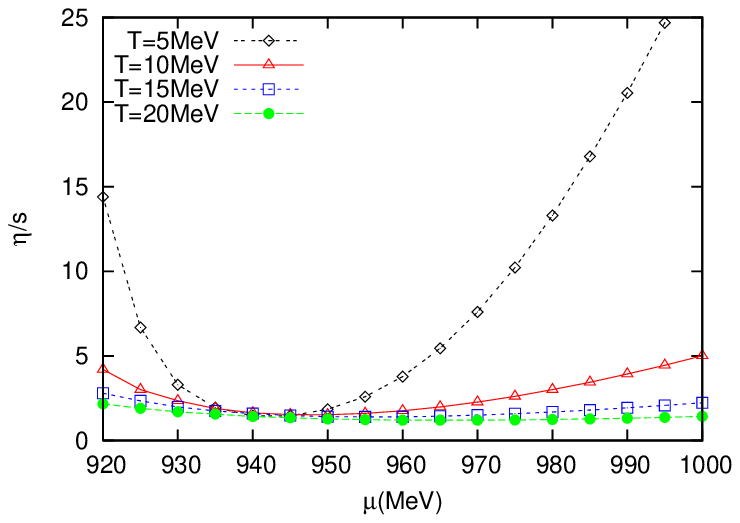}
\hspace{-4mm}
\caption{(Color online) Temperature dependence of the valley structure 
of $\eta/s$ at relatively high chemical potential. Only the 
results at low temperature $T=5,~10,~15,~20$~MeV are shown. 
(Left): Low energy EFT, (Right): Phenomenological amplitudes. } 
\label{valley}
\vspace{-0.6cm}
\end{center}
\end{figure*}

In the right panel of Fig.~\ref{eta/s_mu_dep} 
(low temperature $T=10$~MeV), we pointed out
a valley structure at high baryon chemical potential, 
which was not seen in higher temperature (left panel). 
Let us look at this new structure in more detail and 
examine its possible interpretation. 
We recall again Ref.~\cite{Kapusta}, where it was suggested that 
the ratio $\eta/s$ would give a minimum at the phase transition 
temperature. If this is true for other control parameters, 
in particular, chemical potential, and if there is phase transition 
under the change of chemical potential, it is natural to 
expect that the ratio would exhibit a valley structure 
with its minimum at the critical chemical potential $\mu_c$. 
In other words, if one finds 
a valley structure in the $\mu$ direction, one can expect 
some kind of phase transition around the minimum. This is what 
we observed in Fig.~\ref{eta/s_mu_dep}. Then, what kind of 
phase transition could be related to this valley structure? 
The valley locates at low temperature $T<20$~MeV and 
at high chemical potential $\mu\sim 950$~MeV (which is however 
not enough for the quark-hadron phase transition). 
This is the region where we can see the 
{\it liquid-gas phase transition}. \footnote{Notice that all 
the examples (except for the QCD phase transition) 
discussed in Ref.~\cite{Kapusta} are about the liquid-gas 
phase transitions.} At low $T<20$~MeV and around normal nuclear density, 
there is a critical line (possibly first order) separating 
a nucleon gas phase and a nuclear matter (liquid). This line 
terminates at around $T\sim 15$~MeV \cite{Liquid-gas}, 
and above that temperature, 
there is no distinction between a gas and a liquid. Therefore, 
if the valley indeed corresponds to the liquid-gas phase transition,
it should disappear when the temperature go far beyond $T\sim 15$~MeV. 
In Fig.~\ref{valley}, we show the transition of the valley structure
from $T=5$~MeV up to $T=20$~MeV. The right panel 
is the results of the phenomenological amplitudes. 
Clearly, with increasing temperature, the valley becomes shallow, 
which supports the interpretation that the valley 
structure indeed corresponds to the liquid-gas phase transition.
There are several comments about this. \\

%%%%%%%%%%%%%%%%%%%%%%%%%%%%% figure %%%%%%%%%%%%%%%%%%%%%%%%%%%%%
%%%%%%%%%%%%%%%%%%%%%%%%%%%%% figure %%%%%%%%%%%%%%%%%%%%%%%%%%%%%

%\vspace{-2mm}

\noindent$\bullet$ \ 
 Although the liquid-gas phase transition is a phenomenon
in the hadronic phase, it is not obvious whether the Boltzmann 
equations (valid for a dilute gas) correctly describe the transition 
to liquid phase. Notice that the region where liquid-gas phase 
transition takes place is close to the border of the validity 
region of the Boltzmann equations, and thus it is not surprising 
that there might exist another phase (i.e., liquid phase) outside 
the region of validity. Still, it would be safe to reserve that 
reliable information from our analysis should be only the tendency 
towards phase transition. We do not expect we can 
describe precise structure of phase transition such as the order 
of transition and the temperature dependence of the critical 
chemical potential $\mu_c(T)$ (Our result is that the minimum of 
the valley moves to the right with increasing $T$, 
as opposed to common expectation). In order to describe the transition 
correctly, we will have to include the effects of higher-order 
correlations to the Boltzmann equations, or start from different 
models (such as the $\sigma$-$\omega$ model) which are more 
appropriate for nuclear matter, both of which, however, are 
beyond the scope of the present paper.\\

%\vspace{-2mm}

\noindent$\bullet$ \ 
 As mentioned in Introduction, a similar valley structure 
was already reported in the calculation with the low energy 
EFT \cite{Chen2} and the authors of Ref.~\cite{Chen2}
claim that it is related to the liquid-gas phase transition. 
We have done the same calculation with 
our own parametrizations of low energy EFT, and obtained 
consistent results with Ref.~\cite{Chen2} as shown in 
 the left panel of Fig.~\ref{valley}.
Unlike the results of the 
phenomenological amplitudes (the right panel), 
the valley persists even 
at higher temperature. As we repeated several times, 
since the range of validity of the low energy EFT is severely 
restricted (see the discussion about Fig.~\ref{Average_s_piN}), 
it is quite dangerous 
to draw any conclusions about the physics outside of the validity 
region. However, a similar valley structure is observed even in our 
calculation with the phenomenological amplitudes, and thus our 
calculation partially supports the results of low energy EFT. \\

%\vspace{-2mm}

\noindent$\bullet$ \ 
Since the entropy in our calculation is evaluated by using free 
particle distributions $f_0^{\pi,N}(p)$, there is no information 
about phase transition in the denominator of $\eta/s$. 
However, if there is a real phase transition, entropy will 
of course show nontrivial change around $T_c$ or $\mu_c$. 
For example, in the liquid-gas transition of 
water, both the shear viscosity and the entropy have nontrivial 
structure around $T_c$, and contribute to give a convex shape 
of the ratio $\eta/s$. We emphasize that, in our calculation, 
the valley structure of $\eta/s$ is not the result of entropy. 
If we treated actual entropy, the structure would emerge 
in more pronounced way.

%%%%%%%%%%%%%%%%%%%%%%%%%%%%%%%%%%%%%%%%%%%%%%%%%%%%%
%%%%%%%%%%%%%%%%%%%    Summary   %%%%%%%%%%%%%%%%%%%%
%%%%%%%%%%%%%%%%%%%%%%%%%%%%%%%%%%%%%%%%%%%%%%%%%%%%%

%\setcounter{equation}{0}
\section{Summary}

We have performed a detailed calculation of the shear viscosity 
coefficient $\eta$ and the viscosity to entropy ratio $\eta/s$ 
in a wide region of the hadronic $T$-$\mu$ plane.
Our formalism is based on the relativistic quantum Boltzmann equations,
and we found that it is very important to use phenomenological amplitudes
for the scattering amplitudes in the collision terms in order to 
obtain reliable results. On the other hand, the validity region of 
the low energy effective field theories is severely 
restricted, and reliable results based on them are also limited 
to a small region of $T$-$\mu$ plane. 
We found that the ratio $\eta/s$ decreases (for $T\simge 20$~MeV)
under the inclusion of nucleon degrees of freedom, but still keeps 
above the conjectured KSS bound $\eta/s=1/4\pi$ in the region 
we investigated ($T<180$~MeV, $\mu<1000$~MeV). Since the shear 
viscosity coefficient itself increases with increasing temperature, 
the behavior of the ratio is largely due to the entropy. At low 
temperature $T\simle 15$~MeV and high baryon chemical potential 
$\mu\sim 950$~MeV, we found a valley structure in the ratio. 
There is some argument about the relationship between such 
structure and phase transition, and we expect that 
the valley structure found in our calculation would 
correspond to the liquid-gas phase transition.

\section*{Acknowledgments}
The authors are grateful to E.~Nakano for discussion
and to H.~Hayakawa for explaining them the essence of the 
Chapman-Enskog method. Two of the authors (KI and HO) 
are thankful to S.~Muroya for his encouragements and 
constructive comments. Lastly, one of the authors (HO) thanks 
A.~Dote for useful advises on the fitting method of 
experimental data.\\

%%%%%%%%%%%%%%%%%%%%%%%%%%%%%%%%%
%     appendix
%%%%%%%%%%%%%%%%%%%%%%%%%%%%%%%%%

\appendix
\section{Phenomenological amplitudes}

In this Appendix, we explain the phenomenological amplitudes 
we used in the kinetic equations.  Since the scattering 
amplitude (squared) is related to the differential cross section 
in the CM frame as
\begin{eqnarray}
\left( \frac{d\sigma }{d\Omega } \right)
=\frac{1}{64 \pi^{2} {\sf s}}\, |M|^{2}\,  \, ,
\end{eqnarray}
with ${\sf s}$ being the Mandelstam variable (scattering energy squared),
we discuss only the differential cross sections.

\subsection{$\pi\pi$ scattering}
Consider the partial-wave expansion of the isospin averaged 
$\pi\pi$ elastic differential cross section. Taking up to 
$p$-wave scattering ($\ell =1$) yields a very nice description of the 
experimental data: 
\begin{widetext}
\begin{eqnarray}
\left( \frac{d\sigma }{d\Omega } \right)^{\pi \pi}
&=&\frac{1}{\sum_{I'}(2I'+1)}
\sum_{I=0}^2(2\, I+1) \frac{4}{q_{\rm cm}^{2}} \sum_{\ell=0,1}
\left\vert (2\, \ell+1)\, \e^{i \delta_{\ell}^{I}} 
\sin \delta_{\ell}^{I}\, P_{\ell}(\cos \theta_{\rm cm})  \right\vert^2 \NN
&=&\frac{4}{q_{\rm cm}^{2}}\left(\frac{1}{9}\sin^{2}\delta ^{0}_{0}
+\frac{5}{9}\sin^{2}\delta^{2}_{0}
+\frac{3}{9}\cdot 9\, \sin^{2}\delta^{1}_{1} \cos^{2}\theta_{\rm cm} \right) \, ,
\end{eqnarray}
\end{widetext}
where 4 in the overall factor comes from the identical factor, 
$\theta_{\rm cm}$ and $q_{\rm cm}$ are the scattering angle and 
the magnitude of momentum in the CM frame, 
and $P_{\ell}$ is the $\ell$-th order Legendre polynomial ($P_0(x)=1,\, P_1(x)=x$). 
The numerical factors in the last line are from isospin $I$ 
and orbital angular momentum $\ell$. 
For example,  the last term corresponds to 
$I=\ell=1$ scattering and thus $(3/9)\cdot 9$ is from
the isospin weight $(2I +1)/\sum_{I'}(2I'+1)=3/9$ for $I=1$ 
and $(2\ell+1)^2=9 $ for $\ell=1$.
The phase shift $\delta ^{I}_{\ell}$ depends on $\ell$ and $I$. 
Since the total wavefunction of a $\pi\pi$ system 
must be symmetric under the exchange,
only three phase shifts $(I,\ell)=(0,0),\, (1,1),\, (2,0)$ are possible.
It is known that the energy dependence of the phase shifts are 
well parametrized by the following function \cite{Colangelo:2001df}:
\begin{eqnarray}
&&\hspace{-3mm}\tan\delta ^{I}_{\ell}\NN
&&= \sqrt{1-\frac{4m_{\pi}^{2}}{\sf s}} 
\, q_{\rm cm}^{2\ell}\, \Big(A^{I}_{\ell}
     +B^{I}_{\ell} q_{\rm cm}^2
     +C^{I}_{\ell} q_{\rm cm}^4 
     +D^{I}_{\ell} q_{\rm cm}^6
\Big)\NN
&&\quad \times
\left(\frac{4m_{\pi}^2 -{\sf s}^{I}_{\ell}}
           {{\sf s}-{\sf s}^{I}_{\ell}}
\right)\, ,
\end{eqnarray}   
where the parameters are determined to fit the data up to 
$\sqrt{\sf s}=1.15$ GeV ($q_{\rm cm}\sim$ 550 MeV) and 
are shown in table II:

%%%%%%%%%%%%%%%%%%%%%%%%%%%%% table %%%%%%%%%%%%%%%%%%%%%%%%%%%%%
\begin{table}[h]
\begin{center}
\caption{}
\begin {tabular}{|c|r||c|r||c|r|} \hline
& $(I,\ell)=(0,0)$ & & $(I,\ell)=(1,1)$ & &$(I,\ell)= (2,0)$  \\ \hline
$A^{0}_{0}$ & $2.25\times 10^{-1}$ & $A^{1}_{1}$ & $3.63 \times 10^{-2}$& $A^{2}_{0}$ & $-3.71 \times 10^{-2}$ \\
$B^{0}_{0}$ & $2.46\times 10^{-1}$ & $B^{1}_{1}$ & $1.34 \times 10^{-4}$& $B^{2}_{0}$ & $-8.55 \times 10^{-2}$ \\
$C^{0}_{0}$ & $-1.67 \times 10^{-2}$ & $C^{1}_{1}$ & $-6.98 \times 10^{-5}$& $C^{2}_{0}$ & $-7.54 \times 10^{-3}$ \\
$D^{0}_{0}$ & $-6.40 \times 10^{-4}$ & $D^{1}_{1}$ & $1.41 \times 10^{-6}$& $D^{2}_{0}$ & $-1.99 \times 10^{-4}$ \\ \hline
${\sf s}^{0}_{0}$ & $36.7$ & ${\sf s}^{1}_{1}$ & $30.7$& ${\sf s}^{2}_{0}$ & $-11.9$ \\ \hline
\end{tabular} 
\end{center}
\end{table}
%%%%%%%%%%%%%%%%%%%%%%%%%%%%% table %%%%%%%%%%%%%%%%%%%%%%%%%%%%%

\noindent
In this table, dimensionful parameters are redefined so that the mass 
dimension is provided by $m_\pi$. For example, since $B^0_0$ 
has mass dimension $-2$, we define $B^0_0\equiv b^0_0 /m_\pi^2$
and $b^0_0=2.46\times 10^{-1}$. This phenomenological 
parametrization describes the experimental data 
very well. We use this for the $\pi\pi$ scattering amplitude.

\subsection{$\pi N$ scattering}
Let us now turn to the $\pi N$ scattering. 
Consider the partial-wave expansion of the 
isospin averaged $\pi N$ differential cross section: 
\begin{eqnarray}\label{partial_piN}
\left( \frac{d\sigma }{d\Omega } \right)^{\pi N}\hspace{-5mm}
&=&\!\!\! \frac{1}{q^2_{\rm cm}} \sum^{2\ell_{\rm max}}_{\ell=0}
\left(\frac{2}{6}\, Q^{I=1/2}_{\ell} (q_{\rm cm})
     +\frac{4}{6}\, Q^{I=3/2}_{\ell} (q_{\rm cm}) 
\right)\NN
&&\qquad \times P_{\ell} (\cos\theta )\, .
\end{eqnarray}
The coefficients 
$$
C_\ell (q_{\rm cm})\equiv \frac26Q^{1/2}_{\ell}(q_{\rm cm})
+\frac46 Q^{3/2}_{\ell}(q_{\rm cm})
$$
are functions of $q_{\rm cm}$ and are determined from the 
experimental data. More precisely, each function $C_\ell$ 
is expressed by a superposition of 30 Gaussians: For example, 
for $\ell=0$, we use 
\BQA
C_{\ell=0}(q_{\rm cm})&=&\sum_{n=1}^{15}A_n \exp\left\{
-\left(\frac{q_{\rm cm}-100{\rm MeV}}{60^n {\rm MeV}}\right)^2
\right\}\NN
&+&
\sum_{n=16}^{30}A_n \exp\left\{
-\left(\frac{q_{\rm cm}-800{\rm MeV}}{60^n {\rm MeV}}\right)^2
\right\},\nonumber
\EQA
where positions and widths of Gaussians are found by trial 
and error. We use different values of positions and widths for 
different $\ell$. This kind of technique is sometimes used in 
describing nuclear many body wavefunctions. By using these 
functions and taking the maximum angular momentum $2\ell_{\rm max}=8$, 
we can fit the experimental data \cite{piNdata} up to 
$\sqrt{\sf s}=2$~GeV ($q_{\rm cm}\sim$ 770~MeV). 
Notice that this parametrization works very well at relatively 
high scattering energies, but in fact its quality becomes worse 
at small scattering energies. This is due to the factor 
${1}/{q_{\rm cm}^2}$ in Eq.~(\ref{partial_piN}). If one absorbed 
this factor into the coefficients and performed the same Gaussian 
fitting, quality of the fit would be better even at low scattering 
energies. However, we decided to start from the conventional 
expression shown in Eq.~(\ref{partial_piN}), and to find 
another parametrization for the low energy data. We have 
interpolated the parametrization proposed in Ref.~\cite{EWei} 
which is compactly represented for the scattering amplitude:
\begin{eqnarray}
M_{\pi N}&=&b_{0}
         +b_{1}( \vec{t} \cdot \vec{\tau}  )
         +\Big( c_0
               +c_1 \left( \vec{t}  \cdot \vec{\tau}  \right) 
          \Big)( \vec{q} \cdot \vec{q'})  \NN
         &&+i\Big( d_{0}+d_1\left( \vec{t} \cdot \vec{\tau}  \right)  
           \Big)\, \vec{ \sigma } \cdot (\vec{q}\times \vec{q'})\, ,
\nonumber
\end{eqnarray}
where $\vec{t}$ and $\vec{\tau }/2$ are the isospin vectors of a pion 
and a nucleon respectively, $\vec{q}$ and $\vec{q'}$ are the momenta 
of incoming particles in the CM frame, and lastly 
$\vec{\sigma }$ is the Pauli matrix. The parameters are 
determined as shown in table III (dimensionful parameters are again redefined 
by using $m_\pi$ so that they become dimensionless). 

%%%%%%%%%%%%%%%%%%%%%%%%%%%%% table %%%%%%%%%%%%%%%%%%%%%%%%%%%%%
\begin{table}[h]
\begin{center}
\caption{}
\begin {tabular}{|c|r||c|r|} \hline
$b_{0}$ & $-0.010$ &$b_{1}$ & $-0.091$\\
$c_{0}$ & $0.208$ &$c_{1}$ & $0.175$\\
$d_{0}$ & $-0.190$ &$d_{1}$ & $-0.069$\\ \hline 
\end{tabular} 
\end{center}
\end{table}
%%%%%%%%%%%%%%%%%%%%%%%%%%%%% table %%%%%%%%%%%%%%%%%%%%%%%%%%%%%
%\vspace{-5mm}
\noindent
These two different parametrizations are smoothly matched at 
$\sqrt{\sf s}=1101$ MeV ($q_{\rm cm}$=79 MeV), 
giving a very nice parametrization of the 
experimental data for a wide region of scattering energies.

\subsection{$NN$ scattering}
Let us finally discuss the $NN$ scattering. 
In the text we discussed only the $s$-wave scattering, but here we 
treat arbitrary orbital angular momentum. We define 
the spin-isospin averaged $NN$ differential cross section 
\BQA \label{diff_NN_full}
\left(\frac{d\sigma}{d\Omega}\right)^{NN}_{\rm averaged}
\equiv \frac{1}{16}
&&\hspace{-4mm}\left\{ \left( \frac{d \sigma }{d \Omega } \right)^{0,0}
+3\left( \frac{d \sigma }{d \Omega } \right)^{1,0}\right. \NN
&&\hspace{-7mm}\left. +3\left( \frac{d \sigma }{d \Omega } \right)^{0,1} 
+9\left( \frac{d \sigma }{d \Omega } \right)^{1,1}
\right\} ,
\EQA
where $16=\sum_{I=0,1}(2I+1)\sum_{S=0,1}(2S+1)$ and 
$(d\sigma/d\Omega)^{I,S}$ in the right-hand side 
are the differential cross sections with isospin $I$ and 
spin $S$ specified.
One can further decompose each contribution 
depending on the value of orbital angular momentum $\ell$. 
(Notice that the total $NN$ system must be antisymmetric under the 
exchange of two particles: $(-1)^{\ell+S+I}=-1$.)
\begin{eqnarray}
 \left( \frac{d\sigma }{d\Omega } \right)^{0,0}
&=& \frac{1}{q_{\rm cm}^2}\sum_{\ell={\rm odd}} O^{0,0}_{\ell} (q_{\rm cm})  P_{\ell} (\cos\theta ),\\
 \left( \frac{d\sigma }{d\Omega } \right)^{1,0}
&=& \frac{1}{q_{\rm cm}^2}\sum_{\ell={\rm even}} E^{1,0}_{\ell} (q_{\rm cm})  P_{\ell} (\cos\theta ), \\
 \left( \frac{d\sigma }{d\Omega } \right)^{0,1}
&=& \frac{1}{q_{\rm cm}^2}\sum_{\ell={\rm even}} E^{0,1}_{\ell} (q_{\rm cm})  P_{\ell} (\cos\theta ), \\
 \left( \frac{d\sigma }{d\Omega } \right)^{1,1}
&=& \frac{1}{q_{\rm cm}^2}\sum_{\ell={\rm odd}} O^{1,1}_{\ell} (q_{\rm cm})  P_{\ell} (\cos\theta ).
\end{eqnarray}
Inserting these into Eq.~(\ref{diff_NN_full}), we obtain 
the spin-isospin averaged differential cross section separately 
for even and odd $\ell$:
\begin{widetext}
\BQA
\left(\frac{d\sigma}{d\Omega}\right)^{NN}_{\rm averaged}
=\frac{1}{q_{\rm cm}^2}
      \left[\sum_{\ell={\rm even}}\left\{\frac{3}{16}E^{1,0}_{\ell}
           +\frac{3}{16}E^{0,1}_{\ell}\right\} P_{\ell} (\cos\theta ) 
           +\sum_{\ell={\rm odd}}\left\{\frac{1}{16}O^{0,0}_{\ell}
           +\frac{9}{16}O^{1,1}_{\ell}\right\} P_{\ell} (\cos\theta )
      \right]\, .
\EQA
\end{widetext}
When $\ell=0$, this gives the spin-isospin averaged scattering amplitude
used in the parametrization of low energy effective theory. 

Restriction to the $s$-wave alone may be a good approximation 
at low scattering energy, 
but to obtain a parametrization which describes the data in much wider 
range of energies, we have to include larger $\ell$. To do this, 
we again perform the partial-wave expansion of 
the spin-isospin averaged differential cross section (\ref{diff_NN_full}) and 
fit the energy-dependent coefficients by using the Gaussian 
superposition. We define the coefficients $D_\ell(q_{\rm cm})$ 
similarly to Eq.~(\ref{partial_piN}): 
\begin{eqnarray}\label{partial_NN}
\left( \frac{d\sigma }{d\Omega } \right)^{NN}_{\rm averaged}
= \frac{1}{q_{\rm cm}^2}\sum_{\ell} D_{\ell} (q_{\rm cm})  P_{\ell} (\cos\theta ).
\end{eqnarray}
Since the low energy data are already described well by the 
parametrization (\ref{amp_NN_Swave}) and the parameters given in table I, 
we adopt them as the low energy part of the global parametrization. 
We use them up to $q_{\rm cm}=6.76$ MeV ($\sqrt{\sf s}=1.88$ GeV), 
and beyond that, 
we switch to the partial-wave expansion (\ref{partial_NN}).
For actual fitting of the experimental data \cite{NNonline}, 
we divide the rest of the region into two: 

(i) \ 6.762 MeV $< q_{\rm cm} < 48.76$ MeV \\
{}\qquad\ \, ($1876\, {\rm MeV} <\sqrt{\sf s}<1879\, $MeV)

(ii) 48.76 MeV $< q_{\rm cm} < 405$ MeV \\
{}\qquad\ \, (1879\,MeV $<\sqrt{\sf s}<$ 2043\,MeV)

\noindent
In each region, we determine the coefficient $D_\ell(q_{\rm cm})$ 
similarly as in the case for $\pi N$ scattering, so that the 
curve is smoothly connected at the matching points 
$q_{\rm cm} =6.762$ MeV and $q_{\rm cm}=48.76$ MeV. 
In region (i), the highest value of the angular momentum is taken 
to be 6, while in region (ii), much higher value 16.  
We performed the fit up to $q_{\rm cm} = 405$ MeV 
($\sqrt{\sf s}=$ 2043\,MeV).

\section{Solving the linearized Boltzmann equations}

In this Appendix, we discuss how to solve the Boltzmann equations 
(\ref{bol_pi}) and (\ref{bol_N}). 
First of all, recall that the local equilibrium state is 
defined by the distributions $f_0^{\pi,N}$ which make 
the collision terms (the right hand side of the Boltzmann 
equations) vanishing [see Eqs.~(\ref{eq_pi}), (\ref{eq_N})]. 
However it should be noticed that the local equilibrium is 
{\it not} the solution to the Boltzmann equations: 
Indeed, the parameters $T,\, \mu\, $ and $V^\mu$ characterizing 
$f_0^{\pi,N}$ are dependent on the coordinates, which implies
that the left hand side of the Boltzmann equations (which have 
coordinate derivative $\del_\mu$) do not vanish. 
With this in mind, the Boltzmann equations linearized with 
respect to the deviations $\delta f^{\pi,N}$ from the local 
equilibrium $f_0^{\pi,N}(x,p)$ can read 
as follows (leading order of the left hand side does not contain 
$\delta f^{\pi,N}$):
\begin{widetext}
\begin{eqnarray}
&&\frac{p^{\mu }}{E^{\pi }_{p}}\partial_{\mu}f^{\pi}_{0} 
= {\cal C}^{\pi \pi}[\delta f^{\pi} ,f^{\pi}_{0}] 
 +{\cal C}^{\pi \pi}[f^{\pi}_{0},\delta f^{\pi} ] 
 +{\cal C}^{\pi N} [ \delta f^{\pi},f^{N}_{0} ]  
 +{\cal C}^{\pi N} [ f^{\pi}_{0},\delta f^{N} ]\, ,  \label{chap1} \\
&&\frac{p^{\mu }}{E^{N}_{p}}\partial_{\mu}f^{N}_{0}
={\cal C}^{N \pi}[ \delta f^{N},f^{\pi}_{0} ] 
+{\cal C}^{N \pi}[f^{N}_{0},\delta f^{\pi} ] 
+{\cal C}^{N N} [\delta  f^{N},f^{N}_{0} ]
+{\cal C}^{N N} [ f^{N}_{0},\delta f^{N} ]\, , \label{chap2}
\end{eqnarray}
where the notation for the collision term is for example 
(see Eq.~(\ref{col_pi}))
\BQA
{\cal C}^{\pi \pi}[\delta f^{\pi} ,f^{\pi}_{0}] 
&\equiv& {\cal C}^{\pi \pi}[f^\pi_0+\delta f^{\pi} ,f^{\pi}_{0}] 
-{\cal C}^{\pi \pi}[f_0^{\pi} ,f^{\pi}_{0}] \NN
&=&\frac{g_\pi}{2}\int d\Gamma^{\pi\pi}
\Big\{
\delta f^\pi_{1}f^\pi_{02}(1+ f^\pi_{03})(1+f^\pi_{0p}) 
+ f^\pi_{01}f^\pi_{02}\delta f^\pi_{3}(1+f^\pi_{0p}) 
\NN
&&\hspace{1.6cm}-\delta f^\pi_{1}(1+f^\pi_{02})f^\pi_{03}f^\pi_{0p}
- (1+f^\pi_{01})(1+f^\pi_{02})\delta f^\pi_{3}f^\pi_{0p}
\Big\}\, .
\EQA
This corresponds to the lowest order Chapman-Enskog method. 
So far, the deviations $\delta f^{\pi, N}$ are in principle arbitrary,
but for the purpose of computing the shear viscosity coefficient 
$\eta$, we can restrict only to the deviations that are directly from 
the shear $\del_i V_j\neq 0 \, (i\neq j)$. Thus we ignore any other 
effects except the shear. 

By using the particle number conservation $\del_t n=-n\nabla_iV^i$ 
and the energy-momentum conservation $\del_\nu T^{\mu\nu}=0$, 
%for $f_0^{\pi,N}$ defined by Eqs.~(\ref{eq_pi}) and (\ref{eq_N}), 
the left hand sides 
of Eqs.~(\ref{chap1}) and (\ref{chap2}) can be rewritten as 
\begin{eqnarray}
\frac{p^{\mu }}{E^{\pi }_{p}}\partial_{\mu}f^{\pi}_{ 0}(x) 
&=& \beta\, \frac{f^{\pi}_{0}(1+f^{\pi}_{0})}{E^{\pi}_{p}}
\left(p_{i}p_{j}-\frac{\delta _{ij}}{3} p^{2}\right) 
\left( \nabla ^{i} V^{j} \right)_{\rm trl}\, ,   \label{left_pi} \\
\frac{p^{\mu }}{E^{N}_{p}}\partial_{\mu}f^{N}_{0} (x)
&=&\beta\, \frac{f^{N}_{0}(1-f^{N}_{0})}{E^{N}_{p}}
\left(p_{i}p_{j}-\frac{\delta _{ij}}{3} p^{2}\right) \left( \nabla ^{i} V^{j} \right)_{\rm trl}\, ,\label{left_N} 
\end{eqnarray}
where $\left( \nabla ^{i} V^{j} \right)_{\rm trl}$ is 
defined in Eq.~(\ref{traceless}). Notice that this functional form 
was the motivation for defining new quantities $B^{\pi,N}(p)$
in Eqs.~(\ref{pide}), (\ref{Nde}). Indeed, with respect to 
$B^{\pi,N}(p)$, the right hand sides of Eqs.~(\ref{chap1}) and 
(\ref{chap2}) can be expressed rather compactly.
Introducing further the following notation, 
\begin{eqnarray}
B_{ij}^{\pi,N}(p)\equiv 
\left(\hat{p}_{i} \hat{p}_{j} -\frac{\delta _{ij}}{3}\right)
 B^{\pi,N}(p)\, ,
\end{eqnarray}
the linearized Boltzmann equations (\ref{chap1}) and (\ref{chap2})
can be expressed as ($f^\pi_{01}\equiv f^\pi_0(k_1),\, 
f^\pi_{0p}\equiv f^\pi_0(p)$, etc.)
\begin{eqnarray}
\frac{f^{\pi}_{0p}(1+f^{\pi}_{0p})}{E^{\pi}_{p}}
\left(p_{i}p_{j}-\frac{\delta _{ij}}{3} p^{2}\right) 
&=& \frac{g_{\pi}  }{2} \int d\Gamma^{\pi \pi} 
     (1+f^{\pi}_{01})(1+f^{\pi}_{02})f^{\pi}_{03}f^{\pi}_{0p} 
  \Big(B_{ij}^{\pi}(p)+B_{ij}^{\pi}(k_{3})
          -B_{ij}^{\pi}(k_{2})-B_{ij}^{\pi}(k_{1})  
     \Big)   \NN
&+& g_{N} \!\!\int d\Gamma^{\pi N} 
    (1-f^{N}_{01})(1+f^{\pi}_{02})f^{N}_{03}f^{\pi}_{0p} 
   \Big(B_{ij}^{\pi}(p)+B_{ij}^{N}(k_{3})
         -B_{ij}^{\pi}(k_{2})-B_{ij}^{N}(k_{1})  
    \Big) , \label{chap3}
\end{eqnarray}
\begin{eqnarray}
\frac{f^{N}_{0p}(1-f^{N}_{0p})}{E^{N}_{p}}
\left(p_{i}p_{j}-\frac{\delta_{ij}}{3} p^{2}\right)   
&=&g_{\pi}   \int d\Gamma^{N \pi} 
   (1+f^{\pi}_{01})(1-f^{N}_{02})f^{\pi}_{03}f^{N}_{0p}
 \Big(B_{ij}^{N}(p)+B_{ij}^{\pi}(k_{3})
        -B_{ij}^{N}(k_{2})-B_{ij}^{\pi}(k_{1})  
   \Big) \nonumber  \\
&+&\!\!\frac{g_{N}}{2}\!\! \int\! d\Gamma ^{NN} \!
   (1-f^{N}_{01})(1-f^{N}_{02})f^{N}_{03}f^{N}_{0p} 
   \Big(B_{ij}^{N}(p)+B_{ij}^{N}(k_{3})
        -B_{ij}^{N}(k_{2})-B_{ij}^{N}(k_{1})  
   \Big)\,  .\label{chap4}
\end{eqnarray}
\end{widetext}
These are the equations for $B^{\pi,N}(p)$. 
As explained in the text, we solve these equations by restricting 
$B^{\pi,N}(p)$ to a finite dimensional functional space. 
More precisely, we expand $B^{\pi,N}(p)$ in terms of orthogonal 
polynomial functions $W^{\pi,N}_{(n)}(p)$ [see Eq.~(\ref{expansion})] 
and approximate the series by the first three terms 
[see Eqs.~(\ref{Bpi_exp}) and (\ref{BN_exp})]. 
The expansion coefficients 
($b^{\pi,N}_{(n)}$, $n=0,1,2$) in Eqs.~(\ref{Bpi_exp}) and 
(\ref{BN_exp}) are determined as follows: 
we multiply Eq.~(\ref{chap3}) by 
$\frac{1}{(2\pi)^{3}} (\hat{p_{i}} \hat{p_{j}} -\frac{\delta _{ij}}{3} )
W^{\pi}_{(m)}(p)$ ($m=0,1,2$), 
Eq.~(\ref{chap4}) by 
$\frac{1}{(2\pi)^{3}} (\hat{p_{i}} \hat{p_{j}} -\frac{\delta _{ij}}{3} )
W^{N}_{(m)}(p)$ ($m=0,1,2$), 
and integrate them over $p$. Then, by using the orthogonal conditions 
(\ref{ot1}), (\ref{ot2}),  
we will obtain six independent equations for the coefficients 
$b^{\pi,N}_{(n)}$ ($n=0,1,2$). The resulting equations are still
complicated but can be solved in a numerical way.

%%%%%%%%%%%%%%%%%%%%%%%%%%%%%%%%%%%%%%%%%%%%%%%%%%
%                   REFERENCES
%%%%%%%%%%%%%%%%%%%%%%%%%%%%%%%%%%%%%%%%%%%%%%%%%%

\end{document}